%% file: ms.tex
\documentclass[letterpaper, 10 pt, conference]{ieeeconf} 
\IEEEoverridecommandlockouts   % This command is only needed if you want to use the \thanks command

\makeatletter
\def\endthebibliography{%
  \def\@noitemerr{\@latex@warning{Empty `thebibliography' environment}}%
  \endlist
}
\def\input@path{{./Figures/}{./Sections/}}
\makeatother

\usepackage{cite}
\usepackage{amsmath,amssymb,amsfonts}
\usepackage{algorithmic}
\usepackage{textcomp}
\usepackage{xcolor}
\def\BibTeX{{\rm B\kern-.05em{\sc i\kern-.025em b}\kern-.08em T\kern-.1667em\lower.7ex\hbox{E}\kern-.125emX}}
\usepackage{amssymb}
\let\labelindent\relax % Clash with enumitem
\usepackage{enumitem}
\usepackage{amsthm}% Formats (bold, no indent) theorems, comments and stuff
\newtheoremstyle{dotless}{}{}{\itshape}{}{\bfseries}{}{ }{}
\usepackage{subcaption}
\usepackage{placeins}

\usepackage{amsmath}
\usepackage{graphicx}
\graphicspath{{./Figures/}}
\DeclareGraphicsExtensions{.pdf,.jpeg,.png}

\usepackage{pgfplots}
\pgfplotsset{compat=newest,table/search path={{./Figures/},{./Figures/Model/},{./Figures/Parallel_model_approximator/},{./Figures/Parallel_model_approximator_orthogonality_promoting/}}}
\usepackage{grffile}
\usepackage{ifthen}

\usetikzlibrary{plotmarks}
\usetikzlibrary{arrows.meta}
\usetikzlibrary{patterns,ipe}
\usepgfplotslibrary{patchplots}

\usetikzlibrary{external}
\theoremstyle{dotless}
\newtheorem{dummy}{}
\newtheorem{assumption}[dummy]{Assumption}
\newtheorem{theorem}[dummy]{Theorem}
\newtheorem{corollary}[dummy]{Corollary}
\newtheorem{lemma}[dummy]{Lemma}

\newtheorem{example}[dummy]{Example}
\newtheorem{definition}[dummy]{Definition}
\newtheorem{remark}[dummy]{Remark}

\newcommand{\norm}[2]{\lVert#1\rVert_{#2}}
\newcommand{\abs}[1]{\left\lvert #1 \right\rvert}

\newcommand{\drawlinelegend}[1]{\raisebox{.5ex}{\tikz{\draw[#1, line width=0.4mm] (0,0) -- +(1em, 0);}}}

\newboolean{proofs}
\setboolean{proofs}{true}
\begin{document}

\bstctlcite{IEEEexample:BSTcontrol} % Fix ---- for author names in bibliography

\title{\LARGE \bf Physics-Guided Neural Networks for Feedforward Control: \\ An Orthogonal Projection-Based Approach}

\author{Johan Kon$^1$, Dennis Bruijnen$^2$, Jeroen van de Wijdeven$^3$, Marcel Heertjes$^{1,3}$, and Tom Oomen$^{1,4}$% <-this % stops a space
\thanks{This work is supported by Topconsortia voor Kennis en Innovatie (TKI), and ASML and Philips Engineering Solutions. $^1$: Control Systems Technology Group, Departement of Mechanical Engineering, Eindhoven University of Technology, P.O. Box 513, 5600 MB Eindhoven, The Netherlands, e-mail: j.j.kon@tue.nl. $^2$: Philips Engineering Solutions, High Tech Campus 34, 5656 AE Eindhoven, The Netherlands. $^3$: ASML, De Run 6501, 5504 DR Veldhoven, The Netherlands. $^4$: Delft University of Technology, P.O. Box 5, 2600 AA Delft, The Netherlands. }% <-this % stops a space
}

% m.f.heertjes@tue.nl, t.a.e.oomen@tue.nl.
% e-mail: dennis.bruijnen@philips.com 
\maketitle

\input{Abstract.tex}
 
\input{Introduction.tex}

\input{Problem_formulation.tex}

\input{Non_uniqueness.tex}

\input{Orthogonal_regularization.tex}

\input{Simulation_validation.tex}

\input{Conclusion}

\bibliographystyle{IEEEtran}
\bibliography{IEEEabrv,library.bib}

\end{document}

%% file: Sections/Abstract.tex
\begin{abstract}
Unknown nonlinear dynamics can limit the performance of model-based feedforward control. The aim of this paper is to develop a feedforward control framework for systems with unknown, typically nonlinear, dynamics. To address the unknown dynamics, a physics-based feedforward model is complemented by a neural network. The neural network output in the subspace of the model is penalized through orthogonal projection. This results in uniquely identifiable model coefficients, enabling both increased performance and good generalization. The feedforward control framework is validated on a representative system with performance limiting nonlinear friction characteristics.
\end{abstract}
%The aim of this paper is to develop a feedforward control framework that addresses this 
%The feedforward controller is parametrized as the parallel combination of a model and neural network. 
% that commonly limit performance.

%% file: Sections/Introduction.tex
\section{Introduction}
\label{sec:Introduction}
Feedforward control is a method to significantly improve the performance of dynamic systems \cite{Butterworth2009, Zou2004, Lambrechts2005, Boerlage2003}. In feedforward control, both high performance and task flexibility are desired, i.e., a small tracking error for a variety of references. To realize task flexibility, the feedforward signal is parametrized as the output of a filter driven by a reference \cite{Devasia1996}. It is often desired that this filter is interpretable, i.e., that it has physically meaningful coefficients. Perfect feedforward is guaranteed if the feedforward filter is an accurate description of the inverse dynamics of the system. %Lastly, it is often desired that the filter is interpretable, i.e., that it has physically meaningful coefficients.

Classically, the feedforward controller is parametrized based on physical insights. For example, the inverse dynamics can be explicitly parametrized as a polynomial or rational transfer function \cite{489285}, for which the coefficients can be tuned by hand or learned through data-based methods, e.g., iterative learning control \cite{6837472}. Alternatively, the original system can be identified using system identification tools \cite{sysID_Ljung} and inverted accordingly \cite{VanZundert2018, Tomizuka1987, Devasia2000}. However, this parameterization of the feedforward controller based on physical insights limits the achievable performance in the presence of unknown dynamics \cite{6837472}, and may result in parameter bias.

%The parametrization of the feedforward as a linear system limits the achievable performance in the presence of unknown nonlinear dynamics \cite{6837472}. The nonlinear effects can often not be fully captured by a linear model parametrization, resulting in decreased performance and parameter bias.

To overcome the downsides of the physics-based parametrization in the context of unknown, typically nonlinear dynamics, neural networks have been used in feedforward control as a more flexible model structure to enhance performance \cite{Sorensen1999, HUNT19921083, Otten1997}. Neural networks can provide a rich parametrization of the feedforward controller due to their universal approximator characteristics \cite{Goodfellow2016}, and enable compensation of unknown (nonlinear) dynamics accordingly. However, the dynamics captured in neural networks often lack physical interpretation \cite{Ljung2020} and neural networks are known to extrapolate poorly \cite{Xu2020}.
%outside the training data boundaries\cite{Xu2020}. 
% and are not able to exploit the prior on the dynamic system in the form of a (linear) model.

%Recently, physically meaningful parametrizations in the form of models have been combined with neural networks as universal approximators to combine physical insights and increased performance by learning unmodelled effects \cite{7959606,2017arXiv171011431K}. These physics-guided neural networks (PGNNs) perform better, train faster, and show superior generalization properties when compared to their black-box counterpart due to the inductive bias included with the model. PGNNs have been used in feedforward control in \cite{Bolderman2021}, in which model-based basis functions are used as features to control linear motors. However, these approaches typically do not distinguish between the contributions of the model and the neural network, resulting in limited interpretability and generalization.
% However, little attention has been paid to separate the model from the neural network.

Recently, physically meaningful parametrizations in the form of models have been combined with neural networks as universal approximators to combine physical insights and increased performance by learning unmodelled effects \cite{7959606,2017arXiv171011431K}. These physics-guided neural networks (PGNNs) have been used in feedforward control \cite{Bolderman2021}, in which model-based basis functions are used as features to control linear motors. While these  PGNNs perform better, train faster, and show superior generalization properties when compared to their black-box counterpart, they do not distinguish well between the contributions of the model and the neural network. As a result, the interpretability and generalization are lacking compared to a fully physics-based approach.

%Although some recent works appeared on using neural networks in feedforward control, there does not yet exist a framework in which the contribution of the physical model, capturing the 'known' dynamics, and the neural network, capturing the 'unknown' dynamics, are explicitly separated, such that the physics-based model remains interpretable and \textcolor{red}{provides guarantees} for generalization outside the training regime. The goal of this paper is to develop a feedforward control framework combining physically meaningful models and neural networks to tackle unknown nonlinear dynamics with the aim to increase tracking performance.

Although some recent works appeared on using neural networks in feedforward control, there does not yet exist a framework in which the contribution of the physical model, capturing the known dynamics, and the neural network, capturing the unknown dynamics, are explicitly separated. This paper aims to develop a feedforward control framework that combines a physics-based model with a neural network in such a way that the model remains interpretable and provides guarantees for generalization outside the training regime, while the neural network increases the performance by capturing the unknown dynamics. The main contribution is an orthogonal projection-based cost function to ensure that the neural network compensates only the unknown dynamics.

This contribution consists of the following cornerstones. First, the physics-guided parametrization is introduced in Section \ref{sec:problem_formulation}. Second, in Section \ref{sec:non_uniqueness}, it is shown that this parametrization combined with the least-squares cost function results in an uninterpretable model. Third, the orthogonal projection-based cost function is introduced in Section \ref{sec:uniqueness_through_projection}. Last, in Section \ref{sec:simulation_results}, the developed framework with the projection-based cost function is exemplified by simulation on a system with unknown nonlinear friction characteristics.

\subsection*{Notation and Definitions}
All systems are discrete time, single-input single-output (SISO) with sample rate $f_s$. The sets $\mathbb{Z}_{> 0}$, $\mathbb{R}_{\geq 0}$ represent the set of positive integers and non-negative real numbers respectively. $\textrm{im} \ A$ represents the column space of $A$. 
%i.e., the set of all linear combinations of its columns $A$.  
%and $\mathbb{R}_{+}$ is the set of non-negative rational numbers.
%Signals and vectors of signals: lowercase
%Signals stacked over time window: lowercase with underline
%Time indexing: ()
%Trial indexing: subscript
%Uppercase: matrix
%Caligraphic: systems

%$\textrm{col}$ definitie.
%$\cdotp^*$ is optimal value of optimization.

%% file: Sections/Problem_formulation.tex
\section{Problem Formulation}
\label{sec:problem_formulation}
In this section, feedforward control for dynamic systems is introduced. Then, a physics-guided parametrization based on models and neural networks is introduced for feedforward control. Lastly, the learning problem is formulated.

\subsection{Dynamic Processes and Datasets of Optimal Inputs}
\begin{figure}
\centering
\input{FFW_setup.tex}
%\vspace{-2pt}
\caption{Feedforward setup with input $f$, dynamic system $\mathcal{J}$, reference $r$, and error $e$ (left). The input $f$ is parametrized as the output of a reference dependent filter $\mathcal{F}_{\theta,\phi}$ (right).}
\label{fig:FFW_setup}
\vspace{-15pt}
\end{figure}
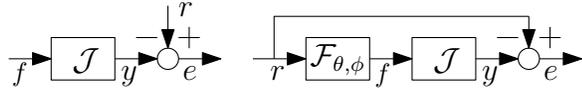

%Feedforward control is a method to a priori compensate  for the dynamic behaviour of a process to track a desired output. 
Consider the feedforward setup in Fig. \ref{fig:FFW_setup} (left). The goal of feedforward control is to ensure that the output $y(k) \in \mathbb{R}$ of a discrete-time (DT) dynamic process $\mathcal{J}$ equals the desired output $r(k) \in \mathbb{R}$ such that the error $e(k) \in \mathbb{R}$, given by
\begin{equation}
	e(k) = r(k) - y(k) = r(k) - \mathcal{J}(f(k)),
\end{equation}
is zero $\forall k \in \mathbb{Z}_{>0}$, in which $f(k) \in \mathbb{R}$ is the input to the process, and $k \in \mathbb{Z}_{>0}$ the time index. The dynamic process $\mathcal{J}$ can represent either a closed-loop or open-loop system, and is defined as follows.
%\textcolor{red}{e.g., the process sensitivity in the linear feedforward setting}. 
\begin{definition}
\label{def:system_class}
The nonlinear dynamic system $\mathcal{J}$ with input $f(k) \in \mathbb{R}$ satisfies the ordinary difference equation (ODE)
%is a non-autonomous ordinary difference equation (ODE) consisting of a linear and nonlinear contribution, given by
% The dynamical system $\mathcal{J}$ is given by the nonlinear ordinary difference equation (ODE)
% \mathcal{J}: f(k) \rightarrow y(k), \ 
\begin{equation}
\begin{aligned}
	{a}^T \tilde{y}(k) + g_y \left( \tilde{y}(k) \right) = f(k),
\end{aligned}
\label{eq:system_class}
\end{equation}
%\tilde{b}^T \tilde{f}(k) + g_f ( \tilde{f}(k) )
with $\tilde{y}(k) = \left[ y(k), \ y^{(1)}(k), \ \hdots, \ y^{(m)}(k) \right]^T \in \mathbb{R}^{m+1}$ the vector of the output and its $m \geq 0$ discrete-time derivatives
\begin{equation}
	y^{(n+1)}(k) = \delta^{n+1} y(k) = f_s \left(y^{(n)}(k) - y^{(n)}(k-1)\right),
\end{equation}
%\ n= 0,\ldots,m-1,
in which $f_s$ is the sampling frequency, and ${a} = \left[ a_0, \ a_1, \ \hdots, \ a_m \right]^T \in \mathbb{R}^{m+1}$ are the coefficients of the linear dynamics of the ODE. $g_y: \mathbb{R} \rightarrow \mathbb{R}$ is an unknown static globally Lipschitz function.
%that can be seen as a perturbation of the linear dynamics.
% $\tilde{b} \in \mathbb{R}^{n+1}$, $\tilde{f}(k) \in \mathbb{R}^{n+1}$, and $g_f$ are analogously defined.
\end{definition}

\begin{remark}
Note that the input $f$ appears linearly, and no derivatives of $f$ appear in the ODE. Consequently, the input $f$ corresponding to a sufficiently smooth desired output $r$ can be found by evaluating \eqref{eq:system_class} for $y=r$.
\end{remark}
%$\mathcal{J}$ consists of a linear part $a^T \tilde{y}(k)$ and an unknown, typically nonlinear, contribution $g_y$.
% that can be seen as a perturbation of the linear dynamics. 
It is assumed that the system is locally nonlinear, as formalized by the following assumption on $g_y$. 
% The dynamic process is approximately linear. 
\begin{assumption}
\label{ass:approximately_linear}
	$g_y$ is zero outside a closed subset $\mathbb{X}$, i.e., 
	\begin{equation}
		g_y \left( \tilde{y}(k) \right) = 0 \quad \forall \tilde{y}(k) \notin \mathbb{X} \subset \mathbb{R}^{m+1}.
	\end{equation}
\end{assumption}
\begin{remark}
\label{rem:LIP_extension}
All results in this paper hold for systems that are linear in parameters (LIP) $a$ with unknown dynamics $g_y$ satisfying Assumption \ref{ass:approximately_linear}. Without loss of generality, the linear model class is considered here as a specific case. 
\end{remark}

For dynamic process $\mathcal{J}$, a dataset $\mathcal{D} = \{ r_j, \hat{f}_j \}_{j=1}^{N_\mathcal{D}}$ of $N_\mathcal{D}$ references $r_j(k)$ with finite length $N_j$ and corresponding optimal inputs $\hat{f}_j(k)$ is available such that
\begin{equation}
	r_j(k) = \mathcal{J}(\hat{f}_j(k)) \ \forall k \in [1, N_j].
\end{equation} 
Note that this implicitly assumes that an input exists that leads to $r_j$. %i.e., that $\mathcal{J}$ is functional controllable. 
This dataset of optimal inputs is used to train a feedforward filter for $\mathcal{J}$ that generates the feedforward $f(k)$ based on reference $r(k)$, i.e., to identify an inverse of $\mathcal{J}$.

% Dit moet netter: gaat erom dat het model globally past op de dynamica, i.e. er is een norm tussen de model input map en de echte input map (integraal over de state space) die bounded is.

\subsection{Physics-Guided Feedforward Parametrization}
To obtain perfect performance, i.e., $e(k) = 0 \ \forall k \in \mathbb{Z}_{>0}$, for a variety of references $r$, the input $f$ is parametrized as the output of a reference-dependent feedforward filter $\mathcal{F}_{\theta,\phi}$ that encapsulates the system class \eqref{eq:system_class}, see Fig. \ref{fig:FFW_setup}. $\mathcal{F}_{\theta,\phi}$ consists of the parallel combination of a physics-based model $\mathcal{M}_\theta$ and a function approximator $\mathcal{C}_\phi$, i.e.,
\begin{equation}
	\mathcal{F}_{\theta,\phi}: r(k) \rightarrow f(k), \ f(k) = \mathcal{M}_\theta (r(k)) + \mathcal{C}_\phi (r(k)).
	\label{eq:feedforward_system}
\end{equation}

%The model allows for explicitly incorporating prior knowledge of the system by a designer. This imposed structure results in interpretability, generalization and sample efficiency. The approximator is a neural network allows for approximating arbitrary functions to compensate for unmodeled non-linear dynamics, thereby allowing for increased performance \cite{7959606,DeGroote2019,Bolderman2021}

\begin{definition}[Model class]
The physics-based feedforward model $\mathcal{M}_\theta$ is parametrized as the weighted sum of the reference and its derivatives to encapsulate the known linear dynamics of \eqref{eq:system_class} according to
\begin{equation}
	\mathcal{M}_\theta: r(k) \rightarrow f_\mathcal{M}(k), \ f_\mathcal{M}(k) = {\theta}^T \tilde{r}(k),
	\label{eq:model_class}
\end{equation}
with parameter vector $\theta = \left[ \theta_0, \ \theta_1, \ \hdots \ \theta_{N_\theta-1} \right]^T \in \mathbb{R}^{N_\theta}$ and 
$\tilde{r}(k) = \begin{bmatrix} r(k) & r^{(1)}(k) & \hdots & r^{(N_\theta-1)}(k) \end{bmatrix}^T \in \mathbb{R}^{N_\theta}$. 
\end{definition}

\begin{definition}[Approximator class]
\label{def:approximator_class}
The approximator $\mathcal{C}_\phi$ is parametrized as a feedforward neural network (FNN) with parameters $\phi$, i.e., 
%\begin{equation}
%	\begin{aligned}
%	\mathcal{C}_\phi: r(k) \rightarrow f_\mathcal{C}(k), \ f_\mathcal{C}(k) =
%	NN(\tilde{r}(k); \phi),
%	\end{aligned}
%\end{equation}
\begin{equation}
\begin{gathered}
	\mathcal{C}_\phi: r(k) \rightarrow f_\mathcal{C}(k), \\ 
	f_\mathcal{C}(k) = W_L \sigma(W_{L-1} \cdots \sigma(W_0 \tilde{r}(k) + b_0) + b_{L-1}),
\end{gathered}
\label{eq:approximator_class}
\end{equation}
%\begin{align}
%	&FNN: \underline{r}(k) \rightarrow f_\mathcal{C}(k) \\ 
%	&f_\mathcal{C}(k) = W_L \sigma(W_{L-1} \cdots \sigma(W_0 \underline{r}(k) + b_0) + b_{L-1}),
%\end{align}
%in which $NN$ is given by
%\begin{equation}
%	f_\mathcal{C}(k) = W_L \sigma(W_{L-1} \cdots \sigma(W_0 \underline{r}(k) + b_0) + b_{L-1})
%\end{equation}
in which $\sigma(\cdotp)$ is an element-wise activation function, such as a sigmoid, hyperbolic tangent, or rectified linear unit (ReLU). The full parameter set of the approximator with $L$ hidden layers is given by $\phi = \{W_i, b_i \}_{i=0}^{L-1} \cup \{ W_L \}$, where $W_i, b_i$ are real, appropriately sized weights and biases. Note that the final layer permits no bias by Assumption \ref{ass:approximately_linear}.
\end{definition}

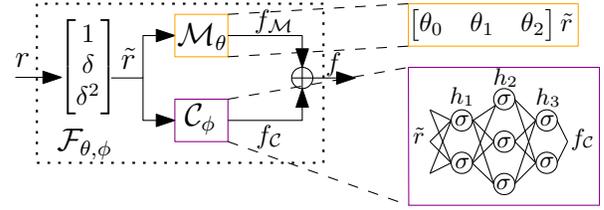
\begin{figure}
\centering
\input{FFW_system3.tex}
%\vspace{-5pt}
\caption{Feedforward filter $\mathcal{F}_{\theta,\phi}$ consisting of model $\mathcal{M}_\theta$ with $N_\theta = 3$, and neural network $\mathcal{C}_\phi$ with $L=3$ hidden layers.}
\label{fig:FFW_filter}
\vspace{-15pt}
\end{figure}

The model structure explicitly incorporates the prior knowledge that $\mathcal{J}$ is linear outside $\mathbb{X}$, i.e., Assumption \ref{ass:approximately_linear}, and is used to extrapolate outside $\mathcal{D}$, whereas the neural network allows for learning the nonlinear dynamics $g_y$. 

For this model and approximator, the feedforward filter is given by (see also Fig. \ref{fig:FFW_filter})
\begin{equation}
	f(k) = \theta^T \tilde{r}(k) + \mathcal{C}_\phi({r}(k)).
	\label{eq:FFW_filter}
\end{equation}

\begin{remark}
	The parametrization of $\mathcal{F}_{\theta,\phi}$ as a parallel linear and nonlinear transformation on $\tilde{r}(k)$ can be interpreted as a single residual layer. Residual layers are favoured over standard layers, as it is easier to approximate deviations from an affine transformation, i.e., the model $\mathcal{M}_\theta$, than to learn an approximate identity mapping over the complete input space \cite{He2016}.
\end{remark}
Given \eqref{eq:system_class} with $m=N_\theta-1$, $\theta = a$ and using $\mathcal{C}_\phi$ to approximate $g_y$, the feedforward filter $\mathcal{F}_{\theta,
\phi}$ is indeed able to model the inverse of $\mathcal{J}$. Thus, it has the potential to generate high performance feedforward with high task flexibility, given correct estimates of $\theta,\phi$.

\subsection{Problem Formulation}
The aim of this paper is to learn the parameters $\theta, \phi$ of $\mathcal{F}_{\theta,\phi}$ in \eqref{eq:FFW_filter}, based on the dataset $\mathcal{D}$, such that $e(k) = 0 \ \forall k \in \mathbb{Z}_{>0}$ for a variety of references $r \notin \mathcal{D}$, i.e., good generalization, while maintaining interpretability of $\mathcal{M}_\theta$. This includes

%a model-aware cost function that recovers a model $\mathcal{M}_\theta$ that is interpretable and representative of the linear dynamics, such that the model can be used for extrapolation and the approximator $\mathcal{C}_\phi$ is truly add-on. 

\begin{enumerate}
	\item illustrating that a standard least squares criterion to fit $\mathcal{F}_{\theta,\phi}$ on $\mathcal{D}$ results in model coefficients $\theta$ that cannot be uniquely determined, resulting in the lack of interpretability and poor generalization of the model,
	\item regularizing the output space of the approximator to recover interpretable model coefficients $\theta$, and
	\item illustrating the proposed approach on a simulated system with Stribeck-like friction characteristics.
\end{enumerate}

%% file: Figures/FFW_setup.tex
\tikzstyle{ipe stylesheet} = [
  ipe import,
  even odd rule,
  line join=round,
  line cap=butt,
  ipe pen normal/.style={line width=0.4},
  ipe pen heavier/.style={line width=0.8},
  ipe pen fat/.style={line width=1.2},
  ipe pen ultrafat/.style={line width=2},
  ipe pen normal,
  ipe mark normal/.style={ipe mark scale=3},
  ipe mark large/.style={ipe mark scale=5},
  ipe mark small/.style={ipe mark scale=2},
  ipe mark tiny/.style={ipe mark scale=1.1},
  ipe mark normal,
  /pgf/arrow keys/.cd,
  ipe arrow normal/.style={scale=7},
  ipe arrow large/.style={scale=10},
  ipe arrow small/.style={scale=5},
  ipe arrow tiny/.style={scale=3},
  ipe arrow normal,
  /tikz/.cd,
  ipe arrows, % update arrows
  <->/.tip = ipe normal,
  ipe dash normal/.style={dash pattern=},
  ipe dash dotted/.style={dash pattern=on 1bp off 3bp},
  ipe dash dashed/.style={dash pattern=on 4bp off 4bp},
  ipe dash dash dotted/.style={dash pattern=on 4bp off 2bp on 1bp off 2bp},
  ipe dash dash dot dotted/.style={dash pattern=on 4bp off 2bp on 1bp off 2bp on 1bp off 2bp},
  ipe dash normal,
  ipe node/.append style={font=\normalsize},
  ipe stretch normal/.style={ipe node stretch=1},
  ipe stretch normal,
  ipe opacity 10/.style={opacity=0.1},
  ipe opacity 30/.style={opacity=0.3},
  ipe opacity 50/.style={opacity=0.5},
  ipe opacity 75/.style={opacity=0.75},
  ipe opacity opaque/.style={opacity=1},
  ipe opacity opaque,
]
\definecolor{red}{rgb}{1,0,0}
\definecolor{blue}{rgb}{0,0,1}
\definecolor{green}{rgb}{0,1,0}
\definecolor{yellow}{rgb}{1,1,0}
\definecolor{orange}{rgb}{1,0.647,0}
\definecolor{gold}{rgb}{1,0.843,0}
\definecolor{purple}{rgb}{0.627,0.125,0.941}
\definecolor{gray}{rgb}{0.745,0.745,0.745}
\definecolor{brown}{rgb}{0.647,0.165,0.165}
\definecolor{navy}{rgb}{0,0,0.502}
\definecolor{pink}{rgb}{1,0.753,0.796}
\definecolor{seagreen}{rgb}{0.18,0.545,0.341}
\definecolor{turquoise}{rgb}{0.251,0.878,0.816}
\definecolor{violet}{rgb}{0.933,0.51,0.933}
\definecolor{darkblue}{rgb}{0,0,0.545}
\definecolor{darkcyan}{rgb}{0,0.545,0.545}
\definecolor{darkgray}{rgb}{0.663,0.663,0.663}
\definecolor{darkgreen}{rgb}{0,0.392,0}
\definecolor{darkmagenta}{rgb}{0.545,0,0.545}
\definecolor{darkorange}{rgb}{1,0.549,0}
\definecolor{darkred}{rgb}{0.545,0,0}
\definecolor{lightblue}{rgb}{0.678,0.847,0.902}
\definecolor{lightcyan}{rgb}{0.878,1,1}
\definecolor{lightgray}{rgb}{0.827,0.827,0.827}
\definecolor{lightgreen}{rgb}{0.565,0.933,0.565}
\definecolor{lightyellow}{rgb}{1,1,0.878}
\definecolor{black}{rgb}{0,0,0}
\definecolor{white}{rgb}{1,1,1}
\begin{tikzpicture}[ipe stylesheet]
  \draw
    (64, 652) rectangle (88, 636);
  \node[ipe node, font=\large]
     at (72, 640) {$\mathcal{J}$};
  \draw
    (108, 644) circle[radius=4];
  \draw[->]
    (108, 664)
     -- (108, 648);
  \node[ipe node, font=\large]
     at (112, 660) {$r$};
  \draw[->]
    (88, 644)
     -- (104, 644);
  \node[ipe node, font=\large]
     at (90.288, 636.208) {$y$};
  \draw[->]
    (112, 644)
     -- (128, 644);
  \draw
    (96, 652)
     -- (104, 652);
  \draw
    (116, 648)
     -- (116, 656);
  \draw
    (112, 652)
     -- (120, 652);
  \node[ipe node, font=\large]
     at (113.644, 636.22) {$e$};
  \draw[->]
    (48, 644)
     -- (64, 644);
  \node[ipe node]
     at (49.207, 635.668) {$f$};
  \draw
    (200, 652) rectangle (224, 636);
  \node[ipe node, font=\large]
     at (208, 640) {$\mathcal{J}$};
  \draw
    (244, 644) circle[radius=4];
  \draw[->]
    (244, 660)
     -- (244, 648);
  \draw[->]
    (224, 644)
     -- (240, 644);
  \node[ipe node, font=\large]
     at (226.288, 636.208) {$y$};
  \draw[->]
    (248, 644)
     -- (264, 644);
  \draw
    (232, 652)
     -- (240, 652);
  \draw
    (252, 648)
     -- (252, 656);
  \draw
    (248, 652)
     -- (256, 652);
  \node[ipe node, font=\large]
     at (249.644, 636.22) {$e$};
  \draw[->]
    (184, 644)
     -- (200, 644);
  \node[ipe node]
     at (186.018, 635.397) {$f$};
  \draw
    (160, 652) rectangle (184, 636);
  \draw[->]
    (140, 644)
     -- (160, 644);
  \node[ipe node, font=\large]
     at (146.649, 636) {$r$};
  \draw
    (148, 644)
     -- (148, 660)
     -- (244, 660)
     -- (244, 660);
  \node[ipe node, font=\large]
     at (161.568, 641.351) {$\mathcal{F}_{\theta,\phi}$};
\end{tikzpicture}

%% file: Figures/FFW_system3.tex
\tikzstyle{ipe stylesheet} = [
  ipe import,
  even odd rule,
  line join=round,
  line cap=butt,
  ipe pen normal/.style={line width=0.4},
  ipe pen heavier/.style={line width=0.8},
  ipe pen fat/.style={line width=1.2},
  ipe pen ultrafat/.style={line width=2},
  ipe pen normal,
  ipe mark normal/.style={ipe mark scale=3},
  ipe mark large/.style={ipe mark scale=5},
  ipe mark small/.style={ipe mark scale=2},
  ipe mark tiny/.style={ipe mark scale=1.1},
  ipe mark normal,
  /pgf/arrow keys/.cd,
  ipe arrow normal/.style={scale=7},
  ipe arrow large/.style={scale=10},
  ipe arrow small/.style={scale=5},
  ipe arrow tiny/.style={scale=3},
  ipe arrow normal,
  /tikz/.cd,
  ipe arrows, % update arrows
  <->/.tip = ipe normal,
  ipe dash normal/.style={dash pattern=},
  ipe dash dotted/.style={dash pattern=on 1bp off 3bp},
  ipe dash dashed/.style={dash pattern=on 4bp off 4bp},
  ipe dash dash dotted/.style={dash pattern=on 4bp off 2bp on 1bp off 2bp},
  ipe dash dash dot dotted/.style={dash pattern=on 4bp off 2bp on 1bp off 2bp on 1bp off 2bp},
  ipe dash normal,
  ipe node/.append style={font=\normalsize},
  ipe stretch normal/.style={ipe node stretch=1},
  ipe stretch normal,
  ipe opacity 10/.style={opacity=0.1},
  ipe opacity 30/.style={opacity=0.3},
  ipe opacity 50/.style={opacity=0.5},
  ipe opacity 75/.style={opacity=0.75},
  ipe opacity opaque/.style={opacity=1},
  ipe opacity opaque,
]
\definecolor{red}{rgb}{1,0,0}
\definecolor{blue}{rgb}{0,0,1}
\definecolor{green}{rgb}{0,1,0}
\definecolor{yellow}{rgb}{1,1,0}
\definecolor{orange}{rgb}{1,0.647,0}
\definecolor{gold}{rgb}{1,0.843,0}
\definecolor{purple}{rgb}{0.627,0.125,0.941}
\definecolor{gray}{rgb}{0.745,0.745,0.745}
\definecolor{brown}{rgb}{0.647,0.165,0.165}
\definecolor{navy}{rgb}{0,0,0.502}
\definecolor{pink}{rgb}{1,0.753,0.796}
\definecolor{seagreen}{rgb}{0.18,0.545,0.341}
\definecolor{turquoise}{rgb}{0.251,0.878,0.816}
\definecolor{violet}{rgb}{0.933,0.51,0.933}
\definecolor{darkblue}{rgb}{0,0,0.545}
\definecolor{darkcyan}{rgb}{0,0.545,0.545}
\definecolor{darkgray}{rgb}{0.663,0.663,0.663}
\definecolor{darkgreen}{rgb}{0,0.392,0}
\definecolor{darkmagenta}{rgb}{0.545,0,0.545}
\definecolor{darkorange}{rgb}{1,0.549,0}
\definecolor{darkred}{rgb}{0.545,0,0}
\definecolor{lightblue}{rgb}{0.678,0.847,0.902}
\definecolor{lightcyan}{rgb}{0.878,1,1}
\definecolor{lightgray}{rgb}{0.827,0.827,0.827}
\definecolor{lightgreen}{rgb}{0.565,0.933,0.565}
\definecolor{lightyellow}{rgb}{1,1,0.878}
\definecolor{black}{rgb}{0,0,0}
\definecolor{white}{rgb}{1,1,1}
\begin{tikzpicture}[ipe stylesheet]
  \draw[->]
    (67.861, 808.2289)
     -- (85.2899, 808.2289);
  \node[ipe node]
     at (84.09, 809.722) {$\begin{bmatrix}
1 \\ \delta \\ \delta^2
\end{bmatrix}$};
  \node[ipe node, font=\large]
     at (67.861, 812.229) {$r$};
  \draw[->]
    (104, 808)
     -- (116, 808)
     -- (116, 824)
     -- (128, 824);
  \draw[->]
    (116, 808)
     -- (116, 792)
     -- (128, 792);
  \draw[orange]
    (128, 832) rectangle (148, 816);
  \node[ipe node, font=\large]
     at (128.796, 820.458) {$\mathcal{M}_\theta$};
  \draw[darkmagenta]
    (128, 800) rectangle (148, 784);
  \node[ipe node, font=\large]
     at (131.771, 789.94) {$\mathcal{C}_\phi$};
  \draw
    (176, 808) circle[radius=4];
  \draw[->]
    (148, 824)
     -- (176, 824)
     -- (176, 812);
  \draw[->]
    (148, 792)
     -- (176, 792)
     -- (176, 804);
  \draw
    (172, 808)
     -- (180, 808);
  \draw
    (176, 812)
     -- (176, 804);
  \draw[->]
    (180, 808)
     -- (196, 808);
  \node[ipe node]
     at (184.873, 810.74) {$f$};
  \node[ipe node]
     at (158.029, 826.852) {$f_\mathcal{M}$};
  \node[ipe node]
     at (158.654, 783.45) {$f_\mathcal{C}$};
  \node[ipe node, font=\large]
     at (107.711, 811.498) {$\tilde{r}$};
  \draw
    (236.1356, 791.9962) circle[radius=4];
  \draw
    (236.1356, 775.9962) circle[radius=4];
  \draw
    (252.1356, 799.9962) circle[radius=4];
  \draw
    (252.1356, 783.9962) circle[radius=4];
  \draw
    (252.1356, 767.9962) circle[radius=4];
  \draw
    (268.1356, 791.9962) circle[radius=4];
  \draw
    (240.1326, 775.8432)
     -- (248.1366, 783.8872);
  \draw
    (240.1316, 775.8162)
     -- (248.1386, 768.1432);
  \draw
    (240.1326, 791.8382)
     -- (248.8616, 797.6982);
  \draw
    (240.4566, 792.0552)
     -- (248.1376, 784.1192);
  \draw
    (256, 768)
     -- (264, 776);
  \draw
    (256.136, 783.996)
     -- (264, 776);
  \draw
    (256, 800)
     -- (264, 776);
  \draw
    (240.1226, 792.3212)
     -- (248.1386, 768.1432);
  \draw
    (240.1256, 776.2812)
     -- (248.8616, 797.6982);
  \draw[darkmagenta]
    (216, 812)
     -- (216.136, 759.996)
     -- (288, 760)
     -- (288, 812)
     -- cycle;
  \draw
    (224.1356, 795.9962)
     -- (232.1356, 775.9962);
  \draw
    (224.1356, 795.9962)
     -- (232.1356, 791.9962);
  \draw
    (224.1356, 783.9962)
     -- (232.1356, 791.9962);
  \draw
    (224.1356, 771.9962)
     -- (232.1356, 791.9962);
  \draw
    (224.1356, 783.9962)
     -- (232.1356, 775.9962);
  \draw
    (224.1356, 771.9962)
     -- (232.1356, 775.9962);
  \draw[ipe pen heavier, ipe dash dotted]
    (76, 836) rectangle (184, 776);
  \node[ipe node, font=\large]
     at (83.565, 780.745) {$\mathcal{F}_{\theta,\phi}$};
  \draw[ipe dash dashed]
    (148, 784)
     -- (216, 760);
  \draw[ipe dash dashed]
    (148, 800)
     -- (216, 812);
  \node[ipe node]
     at (216, 825.627) {$\begin{bmatrix}
\theta_0 & \theta_1 & \theta_2
\end{bmatrix}
\tilde{r}$};
  \draw[orange]
    (216, 836) rectangle (280, 820);
  \draw[ipe dash dashed]
    (148, 832)
     -- (216, 836);
  \draw[ipe dash dashed]
    (148, 816)
     -- (216, 820);
  \draw
    (268.1356, 775.9962) circle[radius=4];
  \draw
    (256, 800)
     -- (264, 792);
  \draw
    (256, 768)
     -- (264, 792);
  \draw
    (256, 784)
     -- (264, 792);
  \draw
    (272, 792)
     -- (276, 784);
  \draw
    (272, 776)
     -- (276, 784);
  \node[ipe node, font=\small]
     at (231.621, 798.387) {$h_1$};
  \node[ipe node, font=\small]
     at (248, 805.518) {$h_2$};
  \node[ipe node, font=\small]
     at (264.031, 798.119) {$h_3$};
  \node[ipe node, font=\small]
     at (277.217, 782.657) {$f_{\mathcal{C}}$};
  \node[ipe node, font=\small]
     at (217.723, 783.399) {$\tilde{r}$};
  \node[ipe node, font=\small]
     at (233.471, 789.945) {$\sigma$};
  \node[ipe node, font=\small]
     at (233.566, 773.913) {$\sigma$};
  \node[ipe node, font=\small]
     at (249.408, 798.103) {$\sigma$};
  \node[ipe node, font=\small]
     at (249.313, 781.976) {$\sigma$};
  \node[ipe node, font=\small]
     at (249.123, 766.229) {$\sigma$};
  \node[ipe node, font=\small]
     at (265.629, 790.039) {$\sigma$};
  \node[ipe node, font=\small]
     at (265.439, 774.292) {$\sigma$};
\end{tikzpicture}

%% file: Sections/Non_uniqueness.tex
\section{Non-Uniqueness and Implications} 
\label{sec:non_uniqueness}
%In this section, it is shown that optimizing $\mathcal{F}_{\theta,\phi}$ according to the commonly employed least-squares cost function does not have a unique optimum for the model coefficients $\theta^*$, resulting in uninterpretable models that do not generalize. 
%which is linked to the notions of identifiability and persistence of excitation.

%\subsection{Least Squares Cost Function}
%With the references and corresponding optimal input in $\mathcal{D}
%$, the parameters of feedforward parametrization $\mathcal{F}_{\theta,\phi}$ are learned, i.e., the parameters of an inverse model set are trained. 
%Fitting the output of $\mathcal{F}_{\theta,\phi}(r(k))$ on $\hat{f}(k)$ is a regression problem, and as such, the least squares criterion is often employed in literature \cite{2017arXiv171011431K, Bolderman2021, HUNT19921083}.

Consider the least-squares criterion $J_{LS} \in \mathbb{R}_{\geq 0}$ given by
%	J_{LS} = \sum_{j=1}^{N_\mathcal{D}} \sum_{k = 1}^{N_j} \abs{\hat{f}_j(k) -  \mathcal{M}_\theta(r_j(k)) - \mathcal{C}_\phi(r_j(k))}^2.
\begin{equation}
	J_{LS} = \sum_{j=1}^{N_\mathcal{D}} \sum_{k = 1}^{N_j} \left(\hat{f}_j(k) -  \mathcal{M}_\theta(r_j(k)) - \mathcal{C}_\phi(r_j(k))\right)^2.
	\label{eq:joint_optimization}
\end{equation}
%However, this naive formulation allows both model $\mathcal{M}_\theta$ and approximator $\mathcal{C}_\phi$ to compensate for the linear dynamics in $\mathcal{J}$ by fitting the same components of the optimal input $\hat{f}_j$, as illustrated in the next example.
This criterion is often employed in literature for regression \cite{2017arXiv171011431K, Bolderman2021, HUNT19921083}. 
% results in uninterpretable models that do not generalize well when used for fitting the output of $\mathcal{F}_{\theta,\phi}(r(k))$ on $\hat{f}(k)$, with $r, \hat{f} \in \mathcal{D}$.
In this section, it is shown that combining $J_{LS}$ with the parallel parametrization $\mathcal{F}_{\theta,\phi}$ in \eqref{eq:FFW_filter} results in an optimum $\arg \min_{\theta,\phi} J_{LS}$ that is not unique, i.e., the model coefficients $\theta^*$ are non-unique. Consequently, the model is uninterpretable. Lastly, it is illustrated that non-unique model coefficients $\theta^*$ prevent generalization.
%results in model coefficients $\theta^*$ that cannot be determined uniquely in combination with $\phi^*$, i.e.,
\subsection{Case 1: Overparametrization of the Feedforward Filter}
One source of non-uniqueness of $\theta^*$ is overparametrization of the parallel filter $\mathcal{F}_{\theta,\phi}$, such that different coefficients $\theta,\phi$ result in the same input-output (IO) behaviour, as formalized in the following definition.
\begin{definition}
\label{def:identifiability}
	The parametrization $\mathcal{F}_{\theta,\phi}$ is identifiable if for two parameter tuples ($\theta_1,\phi_1$), ($\theta_2, \phi_2$) it holds that \cite{sysID_Ljung}
	\begin{equation}
		\mathcal{F}_{\theta_1,\phi_1} = \mathcal{F}_{\theta_2,\phi_2} \Rightarrow (\theta_1,\phi_1) = (\theta_2,\phi_2),
	\end{equation}
	in which the filter equality is defined as
	\begin{equation}
		\mathcal{F}_{\theta_1,\phi_1} = \mathcal{F}_{\theta_2,\phi_2} \Leftrightarrow \mathcal{F}_{\theta_1,\phi_1}(r) = \mathcal{F}_{\theta_2,\phi_2}(r) \ \forall r.
	\end{equation}
\end{definition}
For the naive choice of linear activation functions, $\mathcal{C}_\phi$ reduces to an affine mapping, such that $\mathcal{F}_{\theta,\phi}$ is not identifiable, as formalized by the following result. 
\begin{theorem}
\label{th:identifiability_of_F}
	Given the model class $\mathcal{M}_\theta$ \eqref{eq:model_class} and approximator class $\mathcal{C}_\phi$ \eqref{eq:approximator_class}, the latter with linear activation functions, i.e., $\sigma = I$, then $f_\mathcal{C}(k)$ is an affine map of $\tilde{r}(k)$ according to
	\begin{equation}
		f_\mathcal{C}(k) = \prod_{i=0}^{L} W_i \tilde{r}(k) + \sum_{i=0}^{L-1} b_i \prod_{l=i+1}^{L} W_l = W \tilde{r}(k) + b,
		\label{eq:C_with_linear_activations}
	\end{equation}
	such that the least-squares cost $J_{LS}$  \eqref{eq:joint_optimization} is given by
	\begin{equation}
		\sum_{j=1}^{N_\mathcal{D}} \sum_{k = 1}^{N_j} \abs{\hat{f}_j(k) - \left( (\theta^T + W) \tilde{r}(k) + b\right)}^2.
	\label{eq:joint_optimization_identifiability}
	\end{equation}
\end{theorem}
\ifthenelse{\boolean{proofs}}{%
\begin{proof}
	Substitute $\sigma = I$ in \eqref{eq:approximator_class} and expand the linear map to obtain \eqref{eq:C_with_linear_activations}. Substitute \eqref{eq:C_with_linear_activations} in \eqref{eq:joint_optimization} to obtain \eqref{eq:joint_optimization_identifiability}.
\end{proof}%
}{}%
Theorem \ref{th:identifiability_of_F} indicates that only the sum $\theta^T + W$ can be uniquely determined, i.e., $\mathcal{F}_{\theta,\phi}$ is not identifiable for $\mathcal{C}_\phi$ with linear activation functions. Hence, the optimum $\theta^*, \phi^* = \arg \min_{\theta,\phi} J_{LS}$ is not unique in $\theta^*$ for any dataset $\mathcal{D}$.
%This case corresponds to a feedforward filter that is not identifiable \cite{sysID_Ljung}.
% the input-output (IO) behaviour of the filter is identical for different realizations of its parameters ${\theta, \phi}$
\subsection{Case 2: Persistence of Excitation}
A second source of non-uniqueness of $\theta^*$ is a dataset $\mathcal{D}$ that is not informative enough to distinguish between different coefficients $\theta,\phi$ in $\mathcal{F}_{\theta,\phi}$, as formalized next \cite{sysID_Ljung}.
\begin{definition}
\label{def:persistence_of_excitation}
	A dataset $\mathcal{D}$ is persistently exciting with respect to the identifiable parametrization $\mathcal{F}_{\theta,\phi}$ if, for any two realizations $\mathcal{F}_{\theta_1,\phi_1}, \ \mathcal{F}_{\theta_2,\phi_2}$,
	\begin{equation}
		\sum_{j=1}^{N_\mathcal{D}} \sum_{k=1}^{N_j} \Big(\mathcal{F}_{\theta_1,\phi_1}(r_j(k)) - \mathcal{F}_{\theta_2,\phi_2}(r_j(k)) \Big)^2 = 0,
\end{equation}
implies that $({\theta_1,\phi_1}) =({\theta_2,\phi_2})$.
\end{definition}
When $\mathcal{F}_{\theta,\phi}$ consists of just $\mathcal{M}_\theta$ \eqref{eq:model_class}, conditions on $\mathcal{D}$ to uniquely identify $\theta$ are well-known in terms of the spectrum of $r$ \cite{sysID_Ljung}. However, these results no longer apply to $\mathcal{M}_\theta$ when placed in parallel with $\mathcal{C}_\phi$ in $\mathcal{F}_{\theta,\phi}$, as illustrated next.

%For a non-trivial choice of activation function, $J_{LS}$ in \eqref{eq:joint_optimization} allows both the model $\mathcal{M}_\theta$ and the approximator $\mathcal{C}_\phi$ 
%to fit the known dynamics of $\mathcal{J}$, i.e.,
%to fit the component of the optimal input $\hat{f}_j$ corresponding to the known dynamics of $\mathcal{J}$. In other words, the data
%
\begin{example}
\label{ex:persistence_of_excitation}
Consider $\mathcal{M}_\theta$ \eqref{eq:model_class} of order $N_\theta=2$, and $\mathcal{C}_\phi$ \eqref{eq:approximator_class} with one hidden layer, i.e., $L=1$, with ReLU activation functions, such that $\mathcal{F}_{\theta,\phi}$ is identifiable and given by
\begin{equation*}
	f(k) = \begin{bmatrix} \theta_0 \\ \theta_1 \end{bmatrix}^T \begin{bmatrix} r(k) \\ r^{(1)}(k)\end{bmatrix} + W_1 \max \left(W_0 \begin{bmatrix} r(k) \\ r^{(1)}(k)\end{bmatrix} + b_0 \right).
\end{equation*}
Consider now the dataset $\mathcal{D}$ consisting of a single reference $r$ for which $r(k), r^{(1)}(k) > 0 \ \forall k$, and corresponding optimal input $\hat{f}(k) = c_0 r(k)$, $c_0 \in \mathbb{R}$. Note that $r$ would be persistently exciting for just $\mathcal{M}_\theta$ \cite{sysID_Ljung}. However, the parameters $\theta_0 = c_0 + c_1$, $\theta_1 = 0$, $W_1 = \left[-c_1,\ 0 \right]$, $W_0 = I$ and $b_0 = 0$ lead to $J_{LS} = 0$ for all values of $c_1 \in \mathbb{R}$, as
	\begin{equation*}
		\begin{aligned}
			f(k) &= (c_0 + c_1) r(k) + \begin{bmatrix}
			-c_1 & 0 \end{bmatrix} \max \left( \begin{bmatrix} r(k) \\ r^{(1)}(k)\end{bmatrix} \right)  \\
			&= (c_0 + c_1) r(k) - c_1 r(k) = c_0 r(k) = \hat{f}(k).
		\end{aligned}
	\end{equation*}
Thus the optimum $\theta^*, \phi^* = \arg \min_{\theta,\phi} J_{LS}$ is not unique.
\end{example}%
Similar examples can be given for networks with more layers and different activation functions, illustrating that persistence of excitation conditions on $\mathcal{D}$ associated with linear models $\mathcal{M}_\theta$ are insufficient for a unique estimate $\theta^*$.
%the neural network is always able to also generate an input map that is (locally) linear in $\tilde{r}(k)$, resulting in non-unique model coefficients $\theta$.
%Example \ref{ex:persistence_of_excitation} illustrates that more complex conditions on $\mathcal{D}$ are necesssary to guarantee  
%$J_{LS}$ in \eqref{eq:joint_optimization} allows both the model $\mathcal{M}_\theta$ and the approximator $\mathcal{C}_\phi$ to fit the known dynamics of $\mathcal{J}$, i.e.,
% In this case, the issue is resolved by augmenting the dataset to also include references with negative sign and derivatives.
%
\subsection{Consequences of Non-Uniqueness for Extrapolation}
% \textcolor{red}{Heb het idee dat dit niet scherp genoeg is, en ook te lang, maar ik vind wel dat het hier moet staan (of misschien al claimen in het introducerende stuk van de sectie), omdat de lezer misschien anders ook zoiets heeft van: ja waarom is dit een probleem?}
%While the non-uniqueness of $\theta^*$ does not pose a problem for performance for samples that are similar to the dataset (the IO map around the data is fit correctly in either case)
The non-uniqueness of $\theta^*$ prevents extrapolating outside the dataset based on $\mathcal{M}_\theta$. Under Assumption \ref{ass:approximately_linear}, perfect extrapolation outside $\mathbb{X}$ is guaranteed if
\begin{equation}
	a^T \tilde{y}(k) = f(k) = \theta^T \tilde{r}(k) + \mathcal{C}_\phi({r}(k)),
\end{equation}
i.e., if $\theta = a$ and $\mathcal{C}_\phi({r}(k)) = 0 \ \forall \tilde{r}(k) \notin \mathbb{X}$. This cannot be realized if $\theta^*$ is not unique. Additionally, non-uniqueness can result in uninterpretable models.
%e.g., a physical system with a negative mass estimate. 

Both the problem of identifiability and persistence of excitation are caused by the universal approximator characteristics of $\mathcal{C}_\phi$: the neural network is always able to generate an input map that is (locally) linear in $\tilde{r}(k)$. As a result, both $\mathcal{M}_\theta$ and $\mathcal{C}_\phi$ can capture the known dynamics, resulting in non-unique model coefficients $\theta$.

More rigorous conditions could be imposed on $\mathcal{D}$ to guarantee unique coefficients $\theta$. Instead, to avoid these complex conditions, uniqueness of $\theta^*$ is addressed by modifying the cost $J_{LS}$ such that $\mathcal{M}_\theta$ is prioritized for fitting the known dynamics of $\mathcal{J}$. This modification imposes uniqueness of $\theta^*$ through the cost criterion, essentially favouring one realization of $\mathcal{F}_{\theta,\phi}$ over another based on physical insights.

% Comment Dennis: less risk, predictive maintenance 

%% file: Sections/Orthogonal_regularization.tex
\section{Uniqueness by Orthogonal Regularization}
\label{sec:uniqueness_through_projection}
In this section, the orthogonal projection-based cost function is introduced, which decouples the optimization into orthogonal subspaces and regularizes the output of $\mathcal{C}_\phi$ in the model subspace. As a result, dynamics that can be explained by the model, are explained by the model, resulting in interpretable model coefficients $\theta$ that allow for extrapolation. This constitutes the main contribution of this paper.

\subsection{Model Output Space}
An explicit basis of the output space of model component $\mathcal{M}_\theta$ for reference $r_j$ can be derived by lifting the discrete time signal over the reference length $N_j$. Consider again the model parametrization \eqref{eq:model_class}. For the stacked reference 
\begin{equation}
	\underline{r}_j = \begin{bmatrix} r_j(1) & r_j(2) & \ldots & r_j(N_j) \end{bmatrix}^T \in \mathbb{R}^{N_j},
\end{equation}
the stacked response $\underline{f}_\mathcal{M}$ of $\mathcal{M}_\theta$ is given by
\begin{equation}
\begin{aligned}
	\underline{f}_{\mathcal{M},j} = \begin{bmatrix} \tilde{r}(1) & \ldots \tilde{r}(N_j) \end{bmatrix}^T \theta = M(\underline{r}_j) \theta \in \mathbb{R}^{N_j},
\end{aligned}
\label{eq:LIP_parametrization}
\end{equation}
%\underline{f}^j_\mathcal{M} &= \begin{bmatrix} \tilde{r}^T(1) & \ldots & \tilde{r}^T(N_j) \end{bmatrix} \theta = M(\underline{r}_j) \theta \in \mathbb{R}^{N_j},
%&= \begin{bmatrix} \underline{r}_j & \underline{r}_j^{(1)} & \ldots & \underline{r}_j^{(q)} \end{bmatrix} \theta 
with $M(\underline{r}_j) \in \mathbb{R}^{N_j \times N_\theta}$, $N_j > N_\theta$.
%containing as columns the stacked discrete derivatives of the reference $\underline{r}_j^{(i)},i=0,\ldots,N_\theta - 1$. 
The following is assumed.
\begin{assumption}
	The matrix representation $M(\underline{r}_j)$ of $\mathcal{M}_\theta$ in \eqref{eq:LIP_parametrization} has full rank $N_\theta$ for all references $r_j \in \mathcal{D}$.
	\label{ass:M_full_rank}
\end{assumption}
This corresponds to a persistently exciting dataset $\mathcal{D}$ with respect to the linear model class $\mathcal{M}_\theta$.
%, i.e., the impulse response of the differentiator basis functions of the model.
An explicit basis for the output space of model $\mathcal{M}_\theta$ for reference $r_j$ can be found by the singular value decomposition (SVD) of $M(\underline{r}_j)$.
\begin{definition}
\label{def:SVD}
	The singular value decomposition of full rank matrix $M(\underline{r}_j) \in \mathbb{R}^{N_j \times N_\theta}$,  $N_j > N_\theta$ is the factorization 
	\begin{equation}
		M(\underline{r}_j) = \begin{bmatrix} U_1(\underline{r}_j) & U_2(\underline{r}_j) \end{bmatrix} \begin{bmatrix} \Sigma(\underline{r}_j) \\ 0 \end{bmatrix} V^T(\underline{r}_j),
		\label{eq:SVD}
	\end{equation}
	where $U_1(\underline{r}_j) \in \mathbb{R}^{N_j \times N_\theta}$, $U_2(\underline{r}_j) \in \mathbb{R}^{N_j \times N_j - N_\theta}$ 
	%and $V(\underline{r}_j) \in \mathbb{R}^{N_\theta \times N_\theta}$ 
	%are orthonormal matrices spanning orthogonal subspaces, i.e.,
	satisfy
	% V^T(\underline{r}_j) V(\underline{r}_j) = I_{N_\theta} & & 
%		\begin{aligned}
%		U_1^T(\underline{r}_j) U_1(\underline{r}_j) = \begin{cases} I & \textrm{ if } n=m \\ 0 & \textrm{ if } n \neq m \end{cases}
%	\end{aligned}.
	\begin{equation}
		U_1^T(\underline{r}_j) U_1(\underline{r}_j) = I_{N_\theta},\ U_1^T(\underline{r}_j) U_2(\underline{r}_j) = 0,
	\label{eq:orthonormality_U}
	\end{equation}
	and similarly for $U_2(\underline{r}_j)$.
\end{definition}
%U_2^T(\underline{r}_j) U_2(\underline{r}_j) = I_{N_j - N_\theta} & & U
By Definition \ref{def:SVD}, the response of $\mathcal{M}_\theta$ can be written as
\begin{equation}
	\underline{f}_{\mathcal{M},j} = M(\underline{r}_j) \theta = U_1(\underline{r}_j) \Sigma(\underline{r}_j) V^T(\underline{r}_j) \theta,
\end{equation}
in which the columns of $U_1(\underline{r}_j)$ form a basis for the output space of $\mathcal{M}_\theta$ for reference $r_j$, independent of parameters $\theta$, due to linearity in the parameters of $\mathcal{M}_\theta$.

\subsection{Orthogonal Regularization}
To prevent $\mathcal{C}_\phi$ from learning an input map that could also be represented by $\mathcal{M}_\theta$, i.e., to prioritize the model for fitting the modelled dynamics of $\mathcal{J}$, the stacked output of $\mathcal{C}_\phi$ that lies in the output space of the model spanned by $U_1(\underline{r}_j)$ is penalized through regularization. 

The stacked output of $\mathcal{C}_\phi$ for $\underline{r}_j$ is denoted by
\begin{equation}
	\underline{f}_{\mathcal{C},j} = \begin{bmatrix} \mathcal{C}_\phi\left(\tilde{r}(1)\right) & \ldots & \mathcal{C}_\phi\left(\tilde{r}(N_j)\right) \end{bmatrix}^T = \mathcal{C}_\phi(\underline{r}_j) \in \mathbb{R}^{N_j}.
\end{equation}
The following lemma allows for expressing the component of $\underline{f}_{\mathcal{C},j}$ in the subspace spanned by $U_1(\underline{r}_j)$ \cite{freedman_2009}.
\begin{lemma}
	For models $\mathcal{M}_\theta(r_j)$ that are linear in the parameters with finite-time response $U_1(\underline{r}_j) \Sigma(\underline{r}_j) V^T(\underline{r}_j) \theta$ to reference $\underline{r}_j$, the projection onto the subspace spanned by the columns of $U_1(\underline{r}_j)$ is given by
	\begin{equation}
		\Pi_1(\underline{r}_j) = U_1(\underline{r}_j) U_1^T(\underline{r}_j),
	\end{equation}
	and the projection onto the orthogonal complement spanned by the  columns of $U_2(\underline{r}_j)$ is given by
	\begin{equation}
		\Pi_2(\underline{r}_j) = U_2(\underline{r}_j) U_2^T(\underline{r}_j) = I - U_1(\underline{r}_j) U_1^T(\underline{r}_j).
		\label{eq:complementary_projections}
	\end{equation}
\end{lemma}
%The component of $\underline{f}_{\mathcal{C},j}$ in the subspace spanned by $U_1(\underline{r}_j)$ is thus given by
%%\begin{equation}
%	$\underline{f}_{\mathcal{C},j,1} = \Pi_1(\underline{r}_j) \mathcal{C}(\underline{r}_j)$.
%%\end{equation}
%This component $\underline{f}_{\mathcal{C},j,1}$ is used as orthogonality-promoting regularization to obtain a model-aware cost function.

Thus, the component of $\underline{f}_{\mathcal{C},j}$ in the subspace spanned by $U_1(\underline{r}_j)$ is given by $\Pi_1(\underline{r}_j) \mathcal{C}_\phi(\underline{r}_j)$.
Next, this component is used as regularization to obtain the orthogonal projection-based cost function, constituting the main contribution of this paper.
\begin{definition}
\label{def:model_aware_cost_function}
The cost function $J_{P}\in \mathbb{R}_{\geq 0}$ is defined as the regularized least-squares cost according to
\begin{equation}
	J_{P} = \sum_{j=1}^{N_\mathcal{D}} \left( \norm{\underline{\hat{f}}_j -  \left(M(\underline{r}_j)\theta + \mathcal{C}_\phi(\underline{r}_j) \right)}{2}^2
	 +  \lambda R(\underline{r}_{j}) \right),
	\label{eq:orthogonality_regularized_optimization}
\end{equation}
%+  \lambda R(\underline{f}_{\mathcal{C},j}),
with $\lambda \in \mathbb{R}_{\geq 0}$ the regularization weight, and $R(\underline{r}_{j}): \mathbb{R}^{N_j} \rightarrow \mathbb{R}_{\geq 0}$ the orthogonality-promoting regularization given by
\begin{equation}
	R(\underline{r}_{j}) = \norm{\Pi_1(\underline{r}_j) \mathcal{C}_\phi(\underline{r}_j)}{2}^2.
	\label{eq:orthogonality_promoting_regularization}
\end{equation}
\end{definition}

\begin{remark}
	The regularization $R$ can be interpreted as targeted $L_2$ regularization that only shrinks directions of $\phi$ that generate output in the subspace of the model $\mathcal{M}_\theta$.
\end{remark}

%\begin{remark}
%	The regularization $R$ can be interpreted as targeted $L_2$ regularization that only shrinks coefficients/coefficient directions in $\phi$ that generate output in the subspace of the model.
%\end{remark}

This regularization promotes orthogonality between $\underline{f}_{\mathcal{C},j}$ and \textit{any} $\underline{f}_{\mathcal{M},j}$ through penalizing the component of $\underline{f}_{\mathcal{C},j}$ in the subspace spanned by $U_1(\underline{r}_j)$, due to the linearity in the parameters of $\mathcal{M}_\theta$ \eqref{eq:model_class}. As a result, modelled effects are penalized from being included in the approximator, as formalized in the following theorem.

%\begin{theorem}
%\textcolor{red}{Deze moet anderes, moeite met netjes/strak formuleren}
%	For the given model class $\mathcal{M}_\theta$ \eqref{eq:model_class} and approximator class $\mathcal{C}_\phi$ \eqref{eq:approximator_class}, the optimum $\theta^*, \phi^*$ associated with \eqref{eq:orthogonality_regularized_optimization} is a trade-off between allocating \textcolor{red}{approximation resources} for capturing nonlinear effects to increase performance, and for orthogonality of the model and approximator.
%\end{theorem}

\begin{theorem}
Given the model class $\mathcal{M}_\theta$ \eqref{eq:model_class}, approximator class $\mathcal{C}_\phi$ \eqref{eq:approximator_class} and cost function $J_P$ \eqref{eq:orthogonality_regularized_optimization}, the optimization
\begin{equation}
	\theta^*, \phi^* = \arg \min J_{P}
\end{equation}
can be equivalently written as
\begin{equation}
	\arg \min_{\theta,\phi} J_{P} =  \arg \min_{\theta,\phi} J_1(\theta,\phi) + J_2(\theta,\phi) + J_3(\phi),
\end{equation}
in which $J_i \in \mathbb{R}_{\geq 0}$ are given by
%\begin{align}
%	&\begin{aligned}
%	J_1(\theta,\phi) = \sum_{j=1}^{N_\mathcal{D}} \norm{ & U_{1}(\underline{r}_j) \left( U_{1}^T(\underline{r}_j) \underline{\hat{f}}(\underline{r}_j) \right. - \\ & \left. \Sigma(\underline{r}_j) V^T(\underline{r}_j) \theta - U_{1}^T(\underline{r}_j) \mathcal{C}_{\phi}(\underline{r}_j) \right)}{2}^2,
%	\end{aligned} \label{eq:J_1} \\
%	 &J_2(\theta,\phi) = \sum_{j=1}^{N_\mathcal{D}} \norm{U_{2}(\underline{r}_j) U_{2}^T(\underline{r}_j)\left( \underline{\hat{f}}_{j} - \mathcal{C}_{\phi}(\underline{r}_j) \right)}{2}^2, \label{eq:J_2} \\
%	&J_3(\phi) = \lambda \sum_{j=1}^{N_\mathcal{D}} \norm{U_{1,j}(\underline{r}_j) U_{1}^T(\underline{r}_j) \mathcal{C}_{\phi}(\underline{r}_j)}{2}^2. \label{eq:J_3}
%\end{align}
\begin{align}
	&\begin{aligned}
	J_1(\theta,\phi) = \sum_{j=1}^{N_\mathcal{D}} \norm{ & \Pi_{1}(\underline{r}_j) \underline{\hat{f}}_j - U_{1}(\underline{r}_j) \Sigma(\underline{r}_j) V^T(\underline{r}_j) \theta  \\ 
	- & \Pi_{1}(\underline{r}_j) \mathcal{C}_{\phi}(\underline{r}_j)}{2}^2,
	\end{aligned} \label{eq:J_1} \\
	 &J_2(\theta,\phi) = \sum_{j=1}^{N_\mathcal{D}} \norm{\Pi_{2}(\underline{r}_j)\left( \underline{\hat{f}}_{j} - \mathcal{C}_{\phi}(\underline{r}_j) \right)}{2}^2, \label{eq:J_2} \\
	&J_3(\phi) = \lambda \sum_{j=1}^{N_\mathcal{D}} \norm{\Pi_{1}(\underline{r}_j) \mathcal{C}_{\phi}(\underline{r}_j)}{2}^2. \label{eq:J_3}
\end{align}
\end{theorem}
\ifthenelse{\boolean{proofs}}{%
\begin{proof}
%$U_1(\underline{r}_j), \mathcal{C}_{\phi}(\underline{r}_j)$ are denoted by $U_{1,j}, \mathcal{C}_{\phi,j}$ 
$U_1(\underline{r}_j)$ is denoted by $U_{1,j}$ for brevity (similarly for $\mathcal{C}_\phi, V, \Sigma$). The cost function $J_{P}$ \eqref{eq:orthogonality_regularized_optimization} can be written as
\begin{equation*}
	J_P = \sum_{j=1}^{N_\mathcal{D}} \norm{\underline{\hat{f}}_j - U_{1,j} \Sigma_j V_j^T \theta - \mathcal{C}_{\phi,j}}{2}^2 + \lambda \norm{U_{1,j} U_{1,j}^T \mathcal{C}_{\phi,j}}{2}^2.
\end{equation*}
$\mathcal{C}_{\phi,j}$ and $\underline{\hat{f}}_j$ are split into a contribution in the space spanned by $U_{1,j}$ and by $U_{2,j}$ using \eqref{eq:complementary_projections} as
\begin{equation*}
	\begin{aligned}
	\mathcal{C}_{\phi,j} &= U_{1,j} U_{1,j}^T \mathcal{C}_{\phi,j} + U_{2,j} U_{2,j}^T \mathcal{C}_{\phi,j}, \\
	\underline{\hat{f}}_j &= U_{1,j} U_{1,j}^T \underline{\hat{f}}_{j} + U_{2,j} U_{2,j}^T \underline{\hat{f}}_{j}.
	\end{aligned}
\end{equation*}
%with $U_{1,j} U_{1,j}^T \underline{\hat{f}}_{j}$ the contribution to the optimal input in the space spanned by the columns of  $U_{1,j}$ . 
Consequently, the cost function is split according to
\begin{align*}
J_P =& \sum_{j=1}^{N_\mathcal{D}} \left( \norm{U_{1,j} U_{1,j}^T \underline{\hat{f}}_{j} - U_{1,j} \Sigma_j V_j^T \theta - U_{1,j} U_{1,j}^T \mathcal{C}_{\phi,j} \right. 
\\
&+ \left. U_{2,j} U_{2,j}^T \underline{\hat{f}}_{j} - U_{2,j} U_{2,j}^T \mathcal{C}_{\phi,j}}{2}^2 + \lambda \norm{U_{1,j} U_{1,j}^T \mathcal{C}_{\phi,j}}{2}^2 \right)
\\
=& \sum_{j=1}^{N_\mathcal{D}} \left( \norm{U_{1,j} (U_{1,j}^T \underline{\hat{f}}_{j}  - \Sigma_j V_j^T \theta - U_{1,j}^T \mathcal{C}_{\phi,j})}{2}^2 \right. 
\\ 
&+ \left. \norm{U_{2,j} U_{2,j}^T ( \underline{\hat{f}}_{j} - \mathcal{C}_{\phi,j})}{2}^2 + \lambda \norm{U_{1,j} U_{1,j}^T \mathcal{C}_{\phi,j}}{2}^2 \right),
\end{align*}
in which the equality holds by expanding the norms and using $U^T_{1,j}U_{2,j} = 0$, see Definition \ref{def:SVD}. This expression is equivalent to $\eqref{eq:J_1}$ to $\eqref{eq:J_3}$, thereby completing the proof.
%The first term represents the fit of the optimal input in the output space of the model. Due to the regularization term, the model is preferred to capture this contribution. The second contribution represents the the fit of the optimal input outside the output space of the model. Due to the limited approximation capabilities of a neural network, the optimum is a trade-off between performance orthogonality and 
\end{proof}%
}{}%
\begin{corollary}
If $\mathcal{J}$ is linear, i.e., $g_y(\tilde{y}) = 0 \ \forall \tilde{y}$, and  the model order is greater than or equal to the system order, i.e., $N_\theta \geq m + 1$, then $\underline{\hat{f}}_j \in \text{im } U_1(\underline{r}_j)$, and $\mathcal{C}_\phi = 0$.
\end{corollary}
\ifthenelse{\boolean{proofs}}{%
\begin{proof}
	For this case, $U_{2,j} U_{2,j}^T \underline{\hat{f}}_{j} = 0$ and thus the regularization ensures $\mathcal{C}_{\phi,j} = 0$.
\end{proof}%
}{}%

These contributions create a trade-off between i) capturing $g_y$ with $\mathcal{C}_\phi$ to increase performance ($J_2$) and ii) orthogonality of $\mathcal{M}_\theta$ and $\mathcal{C}_\phi$ ($J_3$). As a result, the modelled dynamics of $\mathcal{J}$ are primarily captured by $\mathcal{M}_\theta$ ($J_1$). 
%Due to the limited approximation capabilities of a neural network, the optimal solution is a trade-off between performance, i.e., picking $\mathcal{C}_\phi$ such that it captures the input due to unmodelled effects $U_{1,j} U_{1,j}^T \underline{\hat{f}}_{2,j}$, and orthogonality, i.e. picking $\mathcal{C}_\phi$ outside the space spanned by $U_{1,j}$.
%  through distributing the limited approximation resources of $\mathcal{C}_\phi$

\begin{remark}
The cost function $J_{LS}$ could recover the same optimum $\theta^*,\phi^*$ as $J_{P}$. However, the regularization \eqref{eq:orthogonality_promoting_regularization} explicitly shapes the cost landscape to recover the orthogonal projection-based solution, whereas for $J_{LS}$ this depends on the initialization of $\phi$, see Section \ref{sec:simulation_results}. 
\end{remark}

The orthogonal projection-based cost function $J_P$ gives direct control over the measure of orthogonality between $\mathcal{M}_\theta$ and $\mathcal{C}_\phi$ through hyperparameter $\lambda$, and can be used to ensure that the modelled dynamics of $\mathcal{J}$ are explained by $\mathcal{M}_\theta$.

% $\lambda = 0$, the naive formulation \eqref{eq:joint_optimization} is recovered. For
% Deze formulatie geeft de user direct controle over de gewenste mate van orthogonaliteit. Voor $\lambda = 0$ krijg je de naive formulation  terug. Voor $\lambda \rightarrow \infty$ krijg je dat de regularizatie een harde constraint wordt.

\subsection{Explicit Orthogonality through Projection}
%The limit of the model-aware cost function \eqref{eq:orthogonality_regularized_optimization} for $\lambda \rightarrow \infty$ corresponds to a solution in which the output of $\mathcal{C}_\phi$ is constrained to the space spanned by $U_2(\underline{r}_j)$. 
For $\lambda \rightarrow\infty$, $J_P$ is equivalent to orthogonality of $\mathcal{M}_\theta$ and $\mathcal{C}_\phi$ through explicit projection of $\mathcal{C}_\phi(\underline{r}_j)$ with $\Pi_2(\underline{r}_j)$. This explicit projection results in a disjoint optimization in which the best linear approximation is recovered for $\mathcal{M}_\theta$, and $\mathcal{C}_\phi$ is fit on the residuals.
% It is shown that for the limit lambda to inf, the optimization can be solved disjoint using any other procedure to identify a model, and fitting a neural network on the residuals.

For $\lambda \rightarrow \infty$, the orthogonality-promoting regularization \eqref{eq:orthogonality_promoting_regularization} asymptotically imposes a constraint, instead of softened regularization, transforming the minimization of \eqref{eq:orthogonality_regularized_optimization} into
\begin{subequations}
	\label{eq:explicit_orthogonality}
	\begin{align}
		J_{P, \infty} = \min_{\theta,\phi} \ & \sum_{j=1}^{N_\mathcal{D}} \norm{\underline{\hat{f}}_j -  \left(M(\underline{r}_j)\theta + \mathcal{C}_\phi(\underline{r}_j) \right)}{2}^2, \label{eq:explicit_orthogonality_cost} \\
		\textrm{s.t. } &	\Pi_1(\underline{r}_j) \mathcal{C}_\phi(\underline{r}_j) = 0,\ j=1,\ldots,N_\mathcal{D}. \label{eq:explicit_orthogonality_constraint}
	\end{align}%
	\label{eq:explicit_orthogonality}%
\end{subequations}%
The constraint \eqref{eq:explicit_orthogonality_constraint} effectively reduces the solution space of $\phi$ for which it holds that $\mathcal{C}_\phi(\underline{r}_j) \in \textrm{im } U_2 (\underline{r}_j)$. An explicit solution to \eqref{eq:explicit_orthogonality_constraint} is to reparametrize $\mathcal{C}_\phi(\underline{r}_j)$ as
\begin{equation}
	\mathcal{C}_\phi (\underline{r}_j) = \Pi_2(\underline{r}_j) \mathcal{Q}_\phi(\underline{r}_j)
	\label{eq:reparametrization_through_projection}
\end{equation}
with $\mathcal{Q}_\phi$ a FNN as in Definition \ref{def:approximator_class}. The full feedforward signal is then given by
\begin{equation}
	\underline{f}_j = M(\underline{r}_j) \theta + \Pi_2(\underline{r}_j) \mathcal{Q}_\phi(\underline{r}_j).
\end{equation}
Note that this requires that $r$ is fully known in advance, and does not allow for realtime adaptation of the reference.

The reparametrization \eqref{eq:reparametrization_through_projection} allows for writing optimization \eqref{eq:explicit_orthogonality} as
\begin{equation}
	J_{P,\infty} = \sum_{j=1}^{N_\mathcal{D}} \norm{\underline{\hat{f}}_j -  \left(M(\underline{r}_j)\theta + \Pi_2(\underline{r}_j)\mathcal{C}_\phi(\underline{r}_j) \right)}{2}^2,
	\label{eq:explicit_orthogonality_projection}
\end{equation}
in which the output of $\mathcal{C}_{\phi}(\underline{r}_j)$ is explicitly constrained to the subspace spanned by $U_2(\underline{r}_j)$ through projection with $\Pi_2(\underline{r}_j)$. This closely relates to linearly constrained neural networks \cite{Hendriks2020}. For the formulation \eqref{eq:explicit_orthogonality_projection}, the following result applies.
\begin{theorem}
\label{th:explicit_orthogonality}
	Given the model class $\mathcal{M}_\theta$ \eqref{eq:model_class} and approximator class $\mathcal{C}_\phi$ \eqref{eq:approximator_class}, then cost function $J_{P,\infty}$ \eqref{eq:explicit_orthogonality_projection} can be written as
	\begin{align}
	J_{P,\infty} =& \min_{\theta} \sum_{j=1}^{N_\mathcal{D}} \norm{  \Pi_1(\underline{r}_j) \underline{\hat{f}}_{j} - U_{1}(\underline{r}_j) \Sigma(\underline{r}_j) V^T(\underline{r}_j) \theta}{2}^2 \nonumber \\ 
	+& \min_{\phi} \sum_{j=1}^{N_\mathcal{D}} \norm{\Pi_{2}(\underline{r}_j) \left( \underline{\hat{f}}_{j} - \mathcal{C}_{\phi}(\underline{r}_j)\right)}{2}^2. \label{eq:explicit_orthogonality_disjoint}
	\end{align}
\end{theorem}
\ifthenelse{\boolean{proofs}}{%
\begin{proof}
 	The cost function $J_{P,\infty}$ in \eqref{eq:explicit_orthogonality_projection} can be written as
 	\begin{equation*}
 		J_{P,\infty} = \min_{\theta,\phi} \sum_{j=1}^{N_\mathcal{D}} \norm{\underline{\hat{f}}_j -  \left(U_{1,j} \Sigma_{j} V_j \theta + U_{2,j} U_{2,j}^T \mathcal{C}_{\phi,j} \right)}{2}^2.
 	\end{equation*}
 	Using Definition \ref{def:SVD}, the optimal input $\underline{\hat{f}}_j$ is decomposed as
\begin{equation*}
	\underline{\hat{f}}_j = U_{1,j} U_{1,j}^T \underline{\hat{f}}_{j} + U_{2,j} U_{2,j}^T \underline{\hat{f}}_{j},
\end{equation*}
with $\underline{\hat{f}}_{1,j} = U_{1,j}^T \underline{\hat{f}}_{j} \in \mathbb{R}^{N_\theta}$ and $\underline{\hat{f}}_{2,j} = U_{2,j}^T \underline{\hat{f}}_{j} \in \mathbb{R}^{N_j - N_\theta}$. Consequently, the optimization is split according to
\begin{equation*}
\begin{aligned}
	\min_{\theta,\phi} \sum_{j=1}^{N_\mathcal{D}} \norm{U_{1,j} \left( \underline{\hat{f}}_{1,j} - \Sigma_j V_j^T \theta \right)
	+ U_{2,j} \left( \underline{\hat{f}}_{2,j} - U_{2,j}^T \mathcal{C}_{\phi,j}\right)}{2}^2,
\end{aligned}
\end{equation*}
or equivalently, by expanding the norm and using $U_{1,j}^T U_{2,j} = 0$, see Definition \ref{def:SVD},
\begin{equation*}
\begin{aligned}
	J_{P,\infty} = &\min_{\theta} \sum_{j=1}^{N_\mathcal{D}} \norm{U_{1,j} \left( U_{1,j}^T \underline{\hat{f}}_{j} - \Sigma_j V_j^T \theta \right)}{2}^2 + \\ &\min_{\phi} \sum_{j=1}^{N_\mathcal{D}} \norm{U_{2,j} \left( U_{2,j}^T\underline{\hat{f}}_{j} - U_{2,j}^T \mathcal{C}_{\phi,j}\right)}{2}^2,
\end{aligned}
\end{equation*}
which is equal to \eqref{eq:explicit_orthogonality_disjoint}, thereby completing the proof.
\end{proof}
}{}%
%Theorem \ref{th:explicit_orthogonality} illustrates that the explicit projection $\Pi_2(\underline{r}_j)\mathcal{C}_\phi(\underline{r}_j)$ results in an optimization disjoint in $\theta$ and $\phi$. This 
%
% implies that a neural network can be combined with a model of \textit{any} model class
%
\begin{remark}
For a linear model, the optimization over $\theta$ in \eqref{eq:explicit_orthogonality_disjoint} corresponds to system identification in the finite impulse response setting. $\theta^*$ then is unique by Assumption \ref{ass:M_full_rank}, and equal to the best linear approximation of $\mathcal{J}$ over $\mathcal{D}$ \cite{Ljung2020}.
\end{remark}
	
	Theorem \ref{th:explicit_orthogonality} illustrates that optimization of $\theta$ is independent of $\phi$ and thus provides a method to add an approximator to a model to improve performance. The add-on characteristic of $\mathcal{C}_\phi = \Pi_2 \mathcal{Q}_\phi$ is then guaranteed by the projection $\Pi_2(\underline{r}_j)$ onto the subspace orthogonal to the model. Additionally, this disjoint optimization implies that $\mathcal{M}_\theta$ can be of any class when obtained beforehand by a different identification method. The downside of this approach is that the unknown dynamics can have a contribution in $U_1(\underline{r}_j)$, such that additional performance is limited by this explicit projection.
	
	% Theorem \ref{th:explicit_orthogonality} provides a method to combine an approximator with \textit{any} model class to improve performance. The optimization in $\theta$ is independent of $\phi$, such that $\mathcal{M}_\theta$ can also be obtained beforehand by any identification method. The add-on characteristics of $\mathcal{C}_\phi$ is then guaranteed by the projection $\Pi_2(\underline{r}_j)$ onto the subspace orthogonal to the fixed model. However, some unknown effects can have a contribution in $U_1(\underline{r}_j)$, such that additional performance is limited by this explicit projection.
	%However, the umodelled effects can bias the parameter estimate $\theta$

% The disjoint optimization resulting from the projection of $\mathcal{C}_\phi(\underline{r}_j)$ onto the subspace spanned by $U_2(\underline{r}_j)$ provides a method to add a neural network to any model to increase performance, in which the model can be obtained beforehand by any identification method. 
%, as will be illustrated in the next section.

% The model-aware regularization only shrinks coefficients that contribute to the input in the model output space, and can thus be interpreted as targeted weight shrinkage.

%% file: Sections/Simulation_validation.tex
\section{Simulation Example}
\label{sec:simulation_results}
In this section, a feedforward filter $\mathcal{F}_{\theta,\phi}$ with cost function $J_{P}$ is demonstrated on an example dynamic system according to Definition \ref{def:system_class}, showing that it outperforms both an approach based solely on $\mathcal{M}_\theta$, as well as an approach based on a parallel combination $\mathcal{F}_{\theta,\phi}$ with criterion $J_{LS}$.

\subsection{Example System}
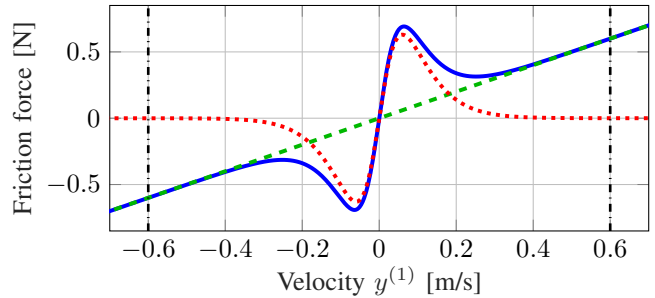
\begin{figure}
\centering
\input{SmoothStribeck.tikz}
%%\vspace{-20pt}
\caption{Stribeck-like friction curve of example system \eqref{eq:stribeck_system} with $c_1 = 1$, $c_2 = 20$, $\alpha = 20$. The nonlinearity $g_y$ (\protect \drawlinelegend{mycolor3,dotted}) is approximately zero outside a closed subset $\mathbb{X}$ (\protect \drawlinelegend{mycolor4,dash dot}). The full friction curve (\protect \drawlinelegend{mycolor1}) coincides with the linear viscous damping (\protect \drawlinelegend{mycolor2,dashed}) for high velocities.}
\label{fig:Stribeck}
%\vspace{-15pt}
\end{figure}

The dynamic process $\mathcal{J}$ is given by a mass-damper system with Stribeck-like friction characteristics, i.e.,
	\begin{equation}
		m y^{(2)}(k) + c_1 y^{(1)}(k) + \frac{c_2 - c_1}{\cosh\left(\alpha y^{(1)}(k)\right)} y^{(1)}(k) = f(k),
		\label{eq:stribeck_system}
	\end{equation}
	with parameters $c_1 = 1$, $c_2 = 20$, $\alpha =20$ and $m=5$, i.e., $a = \left[0,\ 1,\ 5 \right]^T$. The friction characteristics are visualized in Fig. \ref{fig:Stribeck}, in which the nonlinear contribution is negligible outside $\mathbb{X} = \mathbb{R} \times [-0.6, 0.6] \times \mathbb{R}$, thus asymptotically satisfying Assumption \ref{ass:approximately_linear}. For the system \eqref{eq:stribeck_system}, a dataset of $N_\mathcal{D}= 9$ fourth-order references is generated combined with the optimal input $\hat{f}$ for each reference.

\subsection{Simulation Results}
In this section, three feedforward parametrizations and associated optimization criteria are compared. These are 
\begin{enumerate}
	\item A linear model $\mathcal{M}_\theta$ \eqref{eq:model_class} optimized with $J_{LS}$ \eqref{eq:joint_optimization}.
	\item A parallel filter $\mathcal{F}_{\theta,\phi}$ \eqref{eq:feedforward_system} optimized with $J_{LS}$ \eqref{eq:joint_optimization}.
	\item $\mathcal{F}_{\theta,\phi}$ \eqref{eq:feedforward_system} optimized with $J_{P}$ \eqref{eq:orthogonality_regularized_optimization} with $\lambda = 0.01$.
\end{enumerate}

%\definecolor{mycolor1}{rgb}{0.00000,0.44700,0.74100}%
%\definecolor{mycolor2}{rgb}{0.85000,0.32500,0.09800}%
%\definecolor{mycolor3}{rgb}
\definecolor{mycolor1}{rgb}{0.00000,0.72202,0.00000}%
\definecolor{mycolor2}{rgb}{0.00000,0.0,1.00000}%
\definecolor{mycolor3}{rgb}{0.50000,0.0,0.50000}%
\definecolor{mycolor4}{rgb}{0.85000,0.32500,0.09800}%
\definecolor{mycolor5}{rgb}{0.0,0.0,0.0}

\newcommand{\linewidthoptimal}{1.5pt}
\newcommand{\linewidthfilter}{.75pt}
\newcommand{\linewidthmodel}{1.5pt}
\newcommand{\linewidthapproximator}{1.5pt}
\newcommand{\linestyleoptimal}{solid}
\newcommand{\linestylefilter}{solid}
\newcommand{\linestylemodel}{dotted}
\newcommand{\linestyleapproximator}{dashed}
\newcommand{\drawnumber}[3]{\node[draw,circle,minimum size=0.33cm,inner sep=0pt](A) at (axis cs:#1,#2){#3}}
\newcommand{\drawnumbercaption}[1]{\raisebox{-.3ex}{\tikz{\node[draw,circle,minimum size=0.33cm,inner sep=0pt](A) at (0,0){#1}}}}
\newcommand{\circledashstyle}{solid}
%{\raisebox{.5ex}{\draw[#1, line width=0.4mm] (0,0) -- +(1em, 0);}}}
\begin{figure*}[ht]
\centering%
%\begin{subfigure}{.32\textwidth}
%\centering
%\hspace{2em} \textbf{$\boldsymbol{\mathcal{M}_\theta}$ with $\boldsymbol{{J}_{LS}}$}
%\end{subfigure}
%~
%\begin{subfigure}{.32\textwidth}
%\centering
%\hspace{2em} \textbf{$\boldsymbol{\mathcal{F}_{\theta,\phi}}$ with $\boldsymbol{{J}_{LS}}$}
%\end{subfigure}
%~
%\begin{subfigure}{.32\textwidth}
%\centering
%\hspace{2em} \textbf{$\boldsymbol{\mathcal{F}_{\theta,\phi}}$ with $\boldsymbol{{J}_{MA}}$ with $\boldsymbol{\lambda = 0.01}$}
%\end{subfigure}
%~
%\begin{subfigure}[t]{.32\textwidth}
%\input{./Model/ref1.tikz}
%%\vspace{-5pt}
%\end{subfigure}
%~
%\begin{subfigure}[t]{.32\textwidth}
%\input{./Parallel_model_approximator/ref1.tikz}
%%\vspace{-5pt}
%\end{subfigure}
%~
%\begin{subfigure}[t]{.32\textwidth}
%\input{./Parallel_model_approximator_orthogonality_promoting/gamma_0.01/ref1.tikz}
%%\vspace{-5pt}
%\end{subfigure}
%~
\begin{subfigure}[t]{.34\textwidth}%
\input{./Model/ref10.tikz}%
%\vspace{-5pt}
\end{subfigure}%
~%
\begin{subfigure}[t]{.31\textwidth}%
\input{./Parallel_model_approximator/ref10.tikz}%
%\vspace{-5pt}
\end{subfigure}%
~%
\begin{subfigure}[t]{.31\textwidth}%
\input{./Parallel_model_approximator_orthogonality_promoting/ref10.tikz}%
%\vspace{-5pt}
\end{subfigure}%
~%
\\%
\begin{subfigure}[t]{.34\textwidth}%
\input{./Model/ref13.tikz}%
%\vspace{-5pt}
\caption{Physics-based model $\mathcal{M}_\theta$ with $J_{LS}$.}%
\label{fig:col_model}%
\end{subfigure}%
~%
\begin{subfigure}[t]{.31\textwidth}%
\input{./Parallel_model_approximator/ref13.tikz}%
%\vspace{-5pt}
\caption{Parallel combination $\mathcal{F}_{\theta,\phi}$ with $J_{LS}$.}%
%Generated feedforward for a test reference similar to the training references. The approximator has learned a linear input map on top of the nonlinearity. Only when combined with the linear input map of the model, it equals the optimal input . As a result, the obtained model coefficients are not interpretable: the estimated mass $m$ and damping $c_1$ are negative.
%Generated feedforward for a test reference with higher acceleration. Even though the true dynamics of \eqref{eq:stribeck_system} are linear in this dimension, the feedforward filter fails to extrapolate, as the model is not able to capture the true linear dynamics of the system
\label{fig:col_parallel_LS}%
\end{subfigure}%
~%
\begin{subfigure}[t]{.31\textwidth}%
\input{./Parallel_model_approximator_orthogonality_promoting/ref13.tikz}%
%\vspace{-5pt}
\caption{Parallel combination $\mathcal{F}_{\theta,\phi}$ with $J_{P,0.01}$.\label{fig:col_parallel_MA}}%
\end{subfigure}%
\caption{Optimal (\protect \drawlinelegend{mycolor1, \linestyleoptimal}) and generated (\protect \drawlinelegend{mycolor2,\linestylefilter}) feedforward 
%by $\mathcal{M}_\theta$ (left), $\mathcal{F}_{\theta,\phi}$ with cost function $J_{LS}$ (middle) and $\mathcal{F}_{\theta,\phi}$ with cost function $J_{MA}$ (right),
with model component $f_\mathcal{M}$ (\protect \drawlinelegend{mycolor3, \linestylemodel}) and approximator component $f_\mathcal{C}$ (\protect \drawlinelegend{mycolor4, \linestyleapproximator}). 
%The three parametrizations are evaluated for a reference in $\mathcal{D}$ (top), a reference similar to the ones in $\mathcal{D}$ (middle), and a reference dissimilar to the ones in $\mathcal{D}$ (bottom).
The model $\mathcal{M}_\theta$ is not able to capture the unmodelled nonlinearity \protect \drawnumbercaption{1}, and thus has limited performance, but generalizes well due to its linear structure \protect \drawnumbercaption{2}. The parallel combination $\mathcal{F}_{\theta,\phi}$ with cost function $J_{LS}$ is able to capture the unmodelled nonlinear dynamics resulting in high performance for references similar to $\mathcal{D}$ in terms of velocity and acceleration \protect \drawnumbercaption{3}, but has opposing contributions $f_\mathcal{M} \approx - f_\mathcal{C}$ due to non-uniqueness of $\theta^*$, and therefore fails to extrapolate \protect \drawnumbercaption{4}. $\mathcal{F}_{\theta,\phi}$ with orthogonal projection-based cost function $J_{P}$ is both able to capture unmodelled nonlinear dynamics \protect \drawnumbercaption{5}, and to extrapolate based on the modelled dynamics \protect \drawnumbercaption{6}, resulting in high performance, task flexibility, and interpretability.} 
%The parallel filter $\mathcal{F}_{\theta,\phi}$ combined with $J_{P}$ is both able to capture unknown nonlinear effects (top and middle row), and to extrapolate based on the known linear dynamics (bottom row).}%
%In contrast, $\mathcal{M}_\theta$ is not able to capture nonlinear dynamics but extrapolates well. $\mathcal{F}_{\theta,\phi}$ combined with $J_{LS}$ is only able to generate correct input signals for references similar to the ones in the dataset, but extrapolates poorly due to non-unique model coefficients.}
\label{fig:validation}%
\vspace{-15pt}%
\end{figure*}
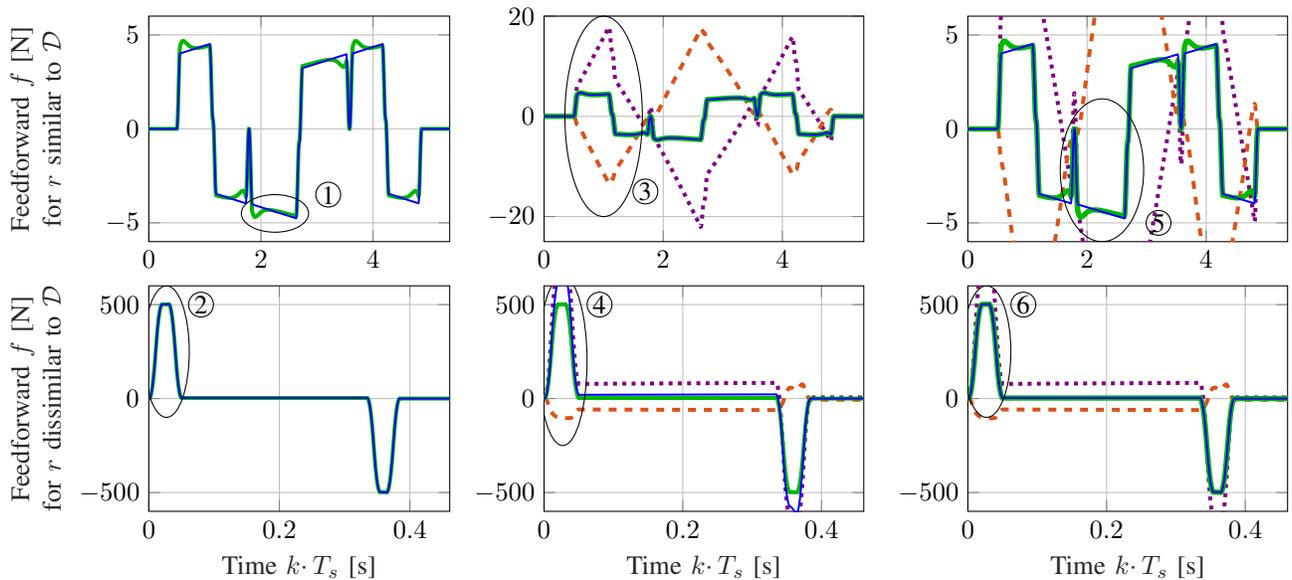%

In all cases, the model $\mathcal{M}_\theta$ is parametrized by a second-order system, i.e., $N_\theta = 3$, according to
	\begin{equation}
		\mathcal{M}_\theta: f_\mathcal{M}(k) = \theta_0 r(k) + \theta_1 r^{(1)}(k) + \theta_2 r^{(2)}(k).
	\end{equation}
	% \theta_0 r(k) +
	In parametrization 2 and 3, $\mathcal{C}_\phi$ is parametrized as a neural network with $L=2$ hidden layers with five neurons each and $\tanh$ activation functions, three input neurons and one linear output neuron, i.e.,
	\begin{equation}
		\mathcal{C}_\phi: f_\mathcal{C}(k) = W_2 \tanh( W_1 \tanh ( W_0 \tilde{r}(k) + b_0) + b_1),
	\end{equation}
	with $\phi = \{W_0,W_1,W_2,b_0,b_1\}$ the parameters of the network, $W_0 \in \mathbb{R}^{5 \times 3}$, $W_1 \in \mathbb{R}^{5 \times 5}$, $W_2 \in \mathbb{R}^{1 \times 5}$, and $\tilde{r}^T(k) = \begin{bmatrix} r(k) & r^{(1)}(k) & r^{(2)}(k) \end{bmatrix}$.
	
Each parametrization is optimized according to the associated criterion with 50 LBFGS iterations, followed by ADAM until convergence. Fig. \ref{fig:validation} shows the generated feedforward signal $f$ by $\mathcal{M}_\theta$ (Fig. \ref{fig:col_model}), $\mathcal{F}_{\theta,\phi}$ with cost function $J_{LS}$ (Fig. \ref{fig:col_parallel_LS}) and $\mathcal{F}_{\theta,\phi}$ with $J_{P}$ (Fig. \ref{fig:col_parallel_MA}), for a reference $r$ similar to $\mathcal{D}$ in terms of position, velocity and acceleration (top), and one dissimilar to $\mathcal{D}$ (bottom) with higher accelerations and velocities, requiring extrapolation outside the training dataset. The following observations are made.
%As a result of the naive optimization $J_{LS}$, the model component of $\mathcal{F}_{\theta,\phi}$ fails to capture the true linear effects of the system: both $\mathcal{M}_\theta$ and $\mathcal{C}_\phi$ contain a linear but opposing contribution to the generated feedforward, that sums to the true linear input map (Fig. \ref{fig:col_parallel_LS}). Even though the performance for references similar to the dataset is good (interpolation), the bad estimate of the true linear behaviour results in 1) uninterpretability of the model, and 2) bad extrapolation for the dominantly linear dynamics. 
%
\begin{itemize}
	\item $\mathcal{M}_\theta$ (Fig. \ref{fig:col_model}) is not able to capture unmodelled nonlinear dynamics in $\hat{f}$: the linear model class limits performance (top). However, specifically due to this prior in the form of a linear model, $\mathcal{M}_\theta$ has satisfactory generalization (bottom) with parameters $\theta^* = \left[0.0,\ 1.18,\ 5.00 \right]^T$.
	%The parameter estimate $\theta^* = \left[0.0,\ 1.18,\ 5.00 \right]^T$ deviates from the true parameters $a$, as it is used to partially capture the behaviour of the nonlinearity over $\mathcal{D}$.
	\item $\mathcal{F}_{\theta,\phi}$ with cost function $J_{LS}$ (Fig. \ref{fig:col_parallel_LS}) is able to capture the nonlinear effects in $\hat{f}$ and thus has good performance for references in and similar to $\mathcal{D}$ (top). However, the model parameters $\theta^* = \left[6.30,\ 25.91,\ 6.90 \right]^T$ are non-unique, not interpretable (no stiffness is present in $\mathcal{J}$), and not equal to the true parameters of the linear dynamics $a$ resulting in bad generalization outside the dataset (bottom).
	\item $\mathcal{F}_{\theta,\phi}$ with cost function $J_{P}$ (Fig. \ref{fig:col_parallel_MA}) is able to capture the unmodelled nonlinear dynamics in $\hat{f}$ with $\mathcal{C}_\phi$ resulting in good performance for references in and similar to $\mathcal{D}$ (top). Additionally, due to the orthogonal projection-based cost function, it has interpretable model coefficients $\theta^* = \left[0,\ 1.17,\ 5.00 \right]$ that accurately describe the true linear dynamics of $\mathcal{J}$ with coefficients $a$, resulting in good generalization (bottom), such that $f_\mathcal{M} \approx \hat{f}$ and $f_\mathcal{C} \approx 0$.
	%The parallel filter $\mathcal{F}_{\theta,\phi}$ (Fig. \ref{fig:col_parallel_MA}) with orthogonal projection-based cost function $J_{P}$ has both good performance and generalization for the linear dynamics, as it is both able to capture the unmodelled nonlinear dynamics (top), and to extrapolate based on the modelled, linear dynamics (bottom row) due to the orthogonal projection-based cost function. 
\end{itemize}
This simulation shows that $\mathcal{F}_{\theta,\phi}$ combined  with $J_{P}$ has both good performance and good generalization, as it is able to capture non-linear effects with $\mathcal{C}_\phi$, and to extrapolate based on the modelled dynamics $\mathcal{M}_\theta$.
%plaatjes met simulatie data met 1) alleen polynomial model, 2) alleen approximator, 3) parallel, 4) parallel met projection, 5) parallel met regularization voor zowel een referentie in de train set als een referentie met extrapolatie. Niet alle opties, alleen parallel lambda = 0 (naive), parallel lambda = normaal (0.01 ofzo), parallel lambda to inf (orthogonal). Wil eigenlijk ook model om te laten zien dat performance slechter is. Ook een approximator om te laten zien dat het niet extrapoleert (dat ook al met lambda = 0). 

%Noemen: alleen model limits performance

%% file: Figures/SmoothStribeck.tikz
\definecolor{mycolor1}{rgb}{0.00000,0.0,1.00000}%
\definecolor{mycolor2}{rgb}{0.00000,0.72202,0.00000}%
\definecolor{mycolor3}{rgb}{1.00000,0.0000,0.00000}%
\definecolor{mycolor4}{rgb}{0.00000,0.0000,0.00000}%

\begin{tikzpicture}

\begin{axis}[%
width=7.1650cm,
height=3cm,
scale only axis,
xmin=-.7,
xmax=.7,
xlabel style={font=\color{white!15!black},yshift=.15cm},
xlabel={Velocity $y^{(1)}$ [m/s]},
ymin=-.85,
ymax=.85,
ylabel style={font=\color{white!15!black}, yshift=-.1cm},
ylabel={Friction force [N]},
axis background/.style={fill=white},
xmajorgrids,
xminorgrids,
ymajorgrids,
yminorgrids
]
\addplot [color=mycolor1, line width=1.5pt, forget plot]
  table[]{SmoothStribeck-1.tsv};
\addplot [color=mycolor2, dashed, line width=1.5pt, forget plot]
  table[]{SmoothStribeck-2.tsv};
\addplot [color=mycolor3, dotted, line width=1.5pt, forget plot]
  table[]{SmoothStribeck-3.tsv};
\addplot [color=mycolor4, dashdotted, line width=1pt, forget plot]
  table[]{SmoothStribeck-4.tsv};
\addplot [color=mycolor4, dashdotted, line width=1pt, forget plot]
  table[]{SmoothStribeck-5.tsv};
\end{axis}
\end{tikzpicture}%

%% file: Figures/Model/ref10.tikz
\begin{tikzpicture}

\begin{axis}[%
width=4.00cm,
height=3cm,
scale only axis,
xmin=0,
xmax=5.366,
xlabel style={font=\color{white!15!black}},
ymin=-6,
ymax=6,
ylabel style={font=\color{white!15!black}, align=center},
ylabel={Feedforward $f$ [N] \\ for $r$ similar to $\mathcal{D}$},
axis background/.style={fill=white},
xmajorgrids,
xminorgrids,
ymajorgrids,
yminorgrids,
every y tick label/.append style={text width=width("$-500$"),align=right}
]
\addplot [color=mycolor1, line width=\linewidthoptimal, \linestyleoptimal]
  table[]{ref10-1.tsv};

\addplot [color=mycolor2, line width=\linewidthfilter, \linestylefilter]
  table[]{ref10-2.tsv};

\draw[\circledashstyle] (axis cs:2.25,-4.5) circle [mycolor5, x radius=.6, y radius=1];

\drawnumber{3.2}{-3.5}{1};

\end{axis}
\end{tikzpicture}%

%% file: Figures/Parallel_model_approximator/ref10.tikz
\begin{tikzpicture}

\begin{axis}[%
width=4.25cm,
height=3cm,
scale only axis,
xmin=0,
xmax=5.366,
xlabel style={font=\color{white!15!black}},
ymin=-25,
ymax=20,
ylabel style={font=\color{white!15!black}, yshift=-.5cm},
axis background/.style={fill=white},
xmajorgrids,
xminorgrids,
ymajorgrids,
yminorgrids,
every y tick label/.append style={text width=width("$-500$"),align=right}
]
\addplot [color=mycolor3, line width=\linewidthmodel, \linestylemodel]
  table[]{../Parallel_model_approximator/ref10-3.tsv};

\addplot [color=mycolor4, line width=\linewidthapproximator, \linestyleapproximator]
  table[]{../Parallel_model_approximator/ref10-4.tsv};

\addplot [color=mycolor1, line width=\linewidthoptimal+.5pt, \linestyleoptimal]
  table[]{../Parallel_model_approximator/ref10-1.tsv};

\addplot [color=mycolor2, line width=\linewidthfilter, \linestylefilter]
  table[]{../Parallel_model_approximator/ref10-2.tsv};

\draw[\circledashstyle] (axis cs:1.0,0) circle [mycolor5, \circledashstyle, x radius=.65, y radius=20];

\drawnumber{1.7}{-15}{3};

\end{axis}
\end{tikzpicture}%

%% file: Figures/Parallel_model_approximator_orthogonality_promoting/ref10.tikz
\begin{tikzpicture}

\begin{axis}[%
width=4.25cm,
height=3cm,
scale only axis,
xmin=0,
xmax=5.366,
xlabel style={font=\color{white!15!black}},
ymin=-6,
ymax=6,
ylabel style={font=\color{white!15!black}},
axis background/.style={fill=white},
xmajorgrids,
xminorgrids,
ymajorgrids,
yminorgrids,
every y tick label/.append style={text width=width("$-500$"),align=right}
]
\addplot [color=mycolor3, line width=\linewidthmodel, \linestylemodel]
  table[]{ref10-3.tsv};

\addplot [color=mycolor4, line width=\linewidthapproximator, \linestyleapproximator]
  table[]{ref10-4.tsv};

\addplot [color=mycolor1, line width=\linewidthoptimal+.5pt, \linestyleoptimal]
  table[]{ref10-1.tsv};

\addplot [color=mycolor2, line width=\linewidthfilter, \linestylefilter]
  table[]{ref10-2.tsv};

\draw[\circledashstyle] (axis cs:2.25,-2.2) circle [mycolor5, x radius=.7, y radius=3.8];

\drawnumber{3.2}{-5}{5};

\end{axis}
\end{tikzpicture}%

%% file: Figures/Model/ref13.tikz
\begin{tikzpicture}

\begin{axis}[%
width=4.00cm,
height=3cm,
scale only axis,
xmin=0,
xmax=0.461,
xlabel style={font=\color{white!15!black}},
xlabel={Time $k \cdotp T_s$ [s]},
ymin=-600,
ymax=600,
ylabel style={font=\color{white!15!black}, align=center},
ylabel={Feedforward $f$ [N] \\ for $r$ dissimilar to $\mathcal{D}$},
axis background/.style={fill=white},
xmajorgrids,
xminorgrids,
ymajorgrids,
yminorgrids,
every y tick label/.append style={text width=width("$-500$"),align=right}
]

\addplot [color=mycolor1, line width=\linewidthoptimal, \linestyleoptimal]
  table[]{../Model/ref13-1.tsv};

\addplot [color=mycolor2, line width=\linewidthfilter, \linestylefilter]
  table[]{../Model/ref13-2.tsv};

\draw[\circledashstyle] (axis cs:0.027,250) circle [mycolor5, x radius=.035, y radius=350];

\drawnumber{0.08}{500}{2};

\end{axis}
\end{tikzpicture}%

%% file: Figures/Parallel_model_approximator/ref13.tikz
\begin{tikzpicture}

\begin{axis}[%
width=4.25cm,
height=3cm,
scale only axis,
xmin=0,
xmax=0.461,
xlabel style={font=\color{white!15!black}},
xlabel={Time $k \cdotp T_s$ [s]},
ymin=-600,
ymax=600,
ylabel style={font=\color{white!15!black}},
axis background/.style={fill=white},
xmajorgrids,
xminorgrids,
ymajorgrids,
yminorgrids,
every y tick label/.append style={text width=width("$-500$"),align=right}
]
\addplot [color=mycolor3, line width=\linewidthmodel, \linestylemodel]
  table[]{../Parallel_model_approximator/ref13-3.tsv};

\addplot [color=mycolor4, line width=\linewidthapproximator, \linestyleapproximator]
  table[]{../Parallel_model_approximator/ref13-4.tsv};

\addplot [color=mycolor1, line width=\linewidthoptimal, \linestyleoptimal]
  table[]{../Parallel_model_approximator/ref13-1.tsv};

\addplot [color=mycolor2, line width=\linewidthfilter, \linestylefilter]
  table[]{../Parallel_model_approximator/ref13-2.tsv};

\draw[\circledashstyle] (axis cs:0.027,200) circle [mycolor5, x radius=.035, y radius=450];

\drawnumber{0.08}{500}{4};

\end{axis}
\end{tikzpicture}%

%% file: Figures/Parallel_model_approximator_orthogonality_promoting/ref13.tikz
\begin{tikzpicture}

\begin{axis}[%
width=4.25cm,
height=3cm,
scale only axis,
xmin=0,
xmax=0.461,
xlabel style={font=\color{white!15!black}},
xlabel={Time $k \cdotp T_s$ [s]},
ymin=-600,
ymax=600,
ylabel style={font=\color{white!15!black}},
axis background/.style={fill=white},
xmajorgrids,
xminorgrids,
ymajorgrids,
yminorgrids,
every y tick label/.append style={text width=width("$-500$"),align=right}
]
\addplot [color=mycolor3, line width=\linewidthmodel, \linestylemodel]
  table[]{ref13-3.tsv};

\addplot [color=mycolor4, line width=\linewidthapproximator, \linestyleapproximator]
  table[]{ref13-4.tsv};

\addplot [color=mycolor1, line width=\linewidthoptimal+.5pt, \linestyleoptimal]
  table[]{ref13-1.tsv};

\addplot [color=mycolor2, line width=\linewidthfilter, \linestylefilter]
  table[]{ref13-2.tsv};

\draw[\circledashstyle] (axis cs:0.027,250) circle [mycolor5, x radius=.035, y radius=350];

\drawnumber{0.08}{500}{6};

\end{axis}
\end{tikzpicture}%

%% file: Sections/Conclusion.tex
\section{Conclusion}
\label{sec:conclusion}
In this paper, a feedforward control framework is introduced that enables superior performance over model-based feedforward control, while maintaining task flexibility and interpretability, for systems that are linear in the parameters with unmodelled, typically nonlinear, dynamics. A physics-based model is complemented by a neural network to arrive at a parallel feedforward filter. For this parallel filter, optimization of its coefficients with a least-squares criterion results in non-unique model coefficients, limiting interpretability and generalization. To address this non-uniqueness, an orthogonal projection-based cost function is derived that regularizes the output of the neural network in the subspace of the model, such that an interpretable model is obtained that accurately captures the modelled system, enabling generalization outside the training dataset. Future work focuses on guaranteeing that the approximator is also regularized to a desired output outside the training dataset, e.g., zero, and extending the approach to flexible dynamics with rational model structures.
% The feedforward framework is validated in simulation on a system with non-linear friction characteristics, illustrating increased performance and generalization.